\documentclass[structabstract]{aa}
\usepackage{graphicx}
\usepackage{txfonts}
\usepackage{natbib}
\usepackage{hyperref}

\begin{document}
   \title{A near-infrared, optical and ultraviolet polarimetric and timing investigation of complex equatorial dusty structures}
   \titlerunning{Polarization of complex dusty structures}
   
   \author{F.~Marin\thanks{\email{frederic.marin@astro.unistra.fr}}
   \and P.~A.~Rojas~Lobos
   \and J.~M.~Hameury
   \and R.~W.~Goosmann}
   
   \authorrunning{Marin et al.}

   \institute{Universit\'e de Strasbourg, CNRS, Observatoire astronomique de Strasbourg, UMR 7550, F-67000 Strasbourg, France}

   \date{Received 15 December, 2017; Accepted 23 January 2018}

  \abstract
  {From stars to active galactic nuclei, many astrophysical systems are 
  surrounded by an equatorial distribution of dusty material that are, 
  in a number of cases, spatially unresolved even with cutting edge facilities.} 
  {In this paper, we investigate if and how one can determine the unresolved 
  and heterogeneous morphology of dust distribution around a central bright 
  source using time-resolved polarimetric observations.}
  {We use polarized radiative transfer simulations to study a sample of 
  circumnuclear dusty morphologies. We explore a grid of uniform, fragmented, 
  density-stratified, geometrically-variable models in the near-infrared, 
  optical and ultraviolet bands, and present their distinctive time-dependent
  polarimetric signatures.} 
  {As expected, varying the structure of the obscuring equatorial disk has a 
  deep impact on the inclination-dependent flux, polarization degree and angle, 
  and time-lags we observe. We find that stratified media are distinguishable 
  by time-resolved polarimetric observations, and that the expected polarization 
  is much higher in the infrared band than in the ultraviolet. However, due to 
  the physical scales imposed by dust sublimation, the average time-lags between 
  the total and polarized fluxes are important (months to years), lengthening 
  the observational campaigns necessary to break more sophisticated (and therefore
  also more degenerated) models. In the ultraviolet band, time-lags are slightly 
  shorter than in the infrared or optical bands, and, coupled to lower diluting 
  starlight fluxes, time-resolved polarimetry in the UV appears more promising 
  for future campaigns.}
  {Equatorial dusty disks differ in terms of inclination-dependent photometric, 
  polarimetric and timing observables, but only the coupling of these different 
  markers can lead to inclination-independent constraints on the unresolved 
  structures. Polarized reverberation mapping in the ultraviolet/blue band, 
  despite being complex and time-consuming, is probably the best technique to 
  rely on in this field.} 


\maketitle

\section{Introduction}
\label{Introduction}
Dusty disks are detected in a variety of astronomical sources that range from stars to the innermost regions of 
active galactic nuclei (AGN). A disk\footnote{Collimated polar outflows, quasi-spherical winds and jets can also 
form during accreting periods but it is beyond the scope of this paper to treat them.} naturally forms when matter 
with sufficient angular momentum is accreted onto a central object, regardless if it is a star, a neutron star or 
a black hole. We know that dust grains can form at the outer rim of AGN and young stars accretion disks 
\citep{Mundy1993,Alessio1998,Alessio1999,Czerny2012} and the fact that disks are ubiquitous in the Universe naturally 
provides clues to the physics of accretion in various environments. By studying dust disks, we can put strong 
constraints on the formation mechanisms of dust grains, and how they survive and evolve. 

Equatorial dusty mixtures can be probed locally, i. e. within the Galaxy, by observing stars. It is thought 
that almost all stars are born with circumstellar disks \citep{Beckwith1996,Hillenbrand1998}. Young stars often 
show protoplanetary disks, which are a collection of material left over from the stellar formation process. 
In Keplerian orbit, this circumstellar, flared region contains both gas and dust (with a gas to dust ratio 
of $\sim$~100, \citealt{Beckwith1990}). While the exact composition of dust in protoplanetary disks is poorly 
known, silicates and amorphous carbons covered by icy mantles are favored (e.g., \citealt{Pollack1994} or, 
more recently, \citealt{Jones2016}). After the accretion stage of the young stellar object, the dust mixture 
in the disk continues to evolve. The innermost grains start to collide and stick together, forming larger grains
that sink towards the disk mid-plane due to the vertical component of gravity. Substantial dust settling can 
profoundly alter the geometry of the protoplanetary disk: the photosphere of the disk is dragged down with the 
dust \citep{Dubrulle1995,Schrapler2004}. Turbulence, if any, tends to mix the grains back up again, resulting 
in non-homogeneous equatorial structures around the star that can be probed in the mid and far-infrared, where 
dust thermally re-emit \citep{Dullemond2007}.

The infrared re-emission of dust was used to infer the existence of dusty disks around more evolved stars, such 
as cool, old white dwarfs in planetary nebulae \citep{Farihi2009,Clayton2014}. A dust-related infrared excess was 
first discovered around the Helix planetary nebula by \citet{Su2007} and later surveys have found several other 
dust-encircled candidates \citep{Chu2011}. The 24~$\mu$m flux densities of hot white dwarfs and central stars of 
planetary nebula is more than two orders of magnitude higher than their expected photospheric emission, revealing 
the presence of cold dust disks \citep{Chu2011}. The disk, in those cases, may derive from the AGB phase of the star
evolution \citep{Clayton2014} or collisions of Kuiper belt-like object \citep{Bilikova2012}. It is not easy to 
distinguish between the two formation mechanisms as the system is not fully resolved. Disk sizes are hard to 
measure because the outer parts are cool and emit weakly, but it is thought that circumstellar disks extend up to 
a few hundreds of astronomical units \citep[see, e.g.,][]{Vicente2005}. The inner part of the dusty disk is set up 
by the sublimation radius that depends on the dust composition and temperature of the central source \citep{Kishimoto2007}. 
Current millimeter and sub-millimeter/interferometers provide images of the dust and gas outer disks with an angular 
resolution better than 1" (i.e. 150~AU at the distance of the nearest star forming regions). As an example, the 
Atacama Large Millimeter/submillimeter Array (ALMA) can reach a resolution of 0.5" at 950~GHz in its most compact 
12-m array configurations. Using longer baselines (up to 14~km), \citet{Andrews2016} successfully traced 
millimeter-sized particles down to spatial scales as small as 1~AU (20~mas). 

High angular resolution observations can also be obtained in the K and N bands, with the Keck Interferometer \citep{Eisner2007}
and the Very Large Telescope Interferometer \citep{Jaffe2004}, respectively. The former authors successfully resolved 
the circumstellar material 2.2~$\mu$m emission within the first astronomical unit around young stars. The later 
spatially resolved a parsec-sized torus-shaped distribution of dust grains in the Seyfert galaxy NGC~1068. This 
circumnuclear dust region was for long invisible to our instruments and is now at the edges of the resolution 
capabilities of current telescopes \citep{Beckert2008}. However, the true morphology and composition of this equatorial 
region is still largely unknown; it would require to resolve sizes inferior to a fraction of a parsec at mega-parsec 
distances. We thus need to find another observation technique to reveal the geometry of unresolvable dusty disks, 
either around young stars or supermassive black holes. 

Indirect techniques such as Doppler tomography \citep{mh88} or eclipse mapping \citep[see e.g.][]{h85} have been 
widely used for cataclysmic variables, but these are restricted to binary systems and cannot be applied to isolated 
stars or AGN. Another very successful method that can go beyond the resolution capabilities of current telescopes 
is the observation of polarized light. Polarization has proven to be independent of the size of the emitting/scattering 
region. Only the observed polarization degree and position angle are sensitive to the morphology, composition and magnetic 
fields of the media that can emit, scatter or absorb photons. Observing the polarized light of NGC~1068, 
\citet{Antonucci1985} successfully unveiled the broad emission lines that were undetectable in the total flux spectrum. 
Those lines, obscured by the equatorial distribution of dust around active galactic nuclei cores, were only 
revealed thanks to near-infrared, optical and ultraviolet spectropolarimetry. They laid the foundations of the 
unified model of AGN \citep{Antonucci1993}. Polarimetry is also extremely powerful in detecting dusty equatorial 
structures around bright stars. The polarized light, emerging from scattering of stellar photons on the disk 
surface, shaves off the unpolarized contribution from the central star. After correcting for stellar starlight 
dilution only the disk emission is detected thanks to its contrast. Imaging polarimetry then allows to reveal 
the disk inner regions down to a few astronomical units \citep[see, e.g.,][]{Gledhill1991,Apai2004}.

Now that we are able to detect the polarized signatures of dusty disks (at least around stars, not yet for AGN),
it is of prime importance to determine their true structure. Models from the literature almost systematically 
use plain, uniform, constant density disks since they are easier to explore numerically. A simple geometry, reducing 
the model parameter space, is much less time-consuming when radiative transfer or magneto-hydrodynamical simulations 
are used. Density-variations have been sometimes included \citep[see, e.g.,][]{Walker2004,Pinte2009} and 
already revealed a number of photometric and spectroscopic differences with respect to constant-density disks. 
The same conclusion applies to the disk geometry \citep{Nenkova2002,Marin2015}. Flared or cylindrical disks 
are the two most common morphologies used in simulations but it has been shown that the structure of 
circumstellar and circumnuclear disks is more complex. Clumpy dusty structures have been detected around stars 
and brown dwarfs (e.g., in the $\sigma$ Ori cluster, see \citealt{Scholz2009}), as well as from galactic structures 
(e.g., in NGC~4244, see \citealt{Holwerda2012}). Even if the parsec-scale dust disk around the core of AGN has not 
yet been resolved to the point of detecting individual clumps, hydrodynamical models predict that large 
homogeneous distributions of dust cannot be stable \citep{Krolik1988}. Various indirect evidences for the clumpiness 
of AGN tori are detected in the X-rays, where fluorescent line emission \citep{Liu2016} and variations in the 
obscuring column density of edge-on Seyferts \citep{Marinucci2016} strongly suggest a non-uniform equatorial 
torus morphology.

It is the aim of this paper to investigate in great details the different morphologies and density stratifications 
of equatorial dusty disks. By doing so, we want to check whether near-infrared, optical and ultraviolet polarimetry 
can shed light on the true geometry of those unresolvable regions. We consider disk parameters (in particular disk 
sizes) appropriate for AGN, but our results can be easily extrapolated to different environments. We thus study 
very realistic models of dusty environments that exist in accreting objects and present the radiative transfer code 
we used in Sect.~\ref{Model}. We compile and discuss our results in Sect.~\ref{Results}, where the photometric flux, 
polarization degree, polarization angle and time-lags for each of the fifteen models are presented. We 
analyze the benefits, problems and the limitations of polarized time reverberation studies in Sect.~\ref{Discussion} 
by comparing our conclusions to recent observations, and we conclude our paper in Sect.~\ref{Conclusions}.

\section{Models of equatorial dusty structures}
\label{Model}
In the following, we use for a geometrical configuration that matches the observational constraints for AGN. 
Our goal is not to precisely model a circumstellar or a circumnuclear disk. We instead test the impact of different 
geometries and compositions onto the resulting photometric flux, polarization and time-lags. We consider throughout 
this paper a flared disk with inner radius 0.1~pc and outer radius 10~pc. The first value corresponds to the inner
radius set by dust sublimation theories for a typical Seyfert galaxy whose dust is mainly composed of graphite and 
silicate grains \citep{Kishimoto2007}. The outer radius is the typical torus extension set by modeling constraints 
based on spectroscopic and interferometric observations and is of order of 5 -- 10~pc 
\citep{Meisenheimer2008,Burtscher2013,Netzer2015,Lopez2016}. The half-opening angle of the flared disk is set to 
45$^\circ$. This corresponds to the estimated torus height derived from quasar observations \citep{Sazonov2015}, 
numerical modeling \citep{Marin2012,Gnedin2015}, and statistical analyses \citep{Marin2014b,Marin2016}. 

\subsection{Dusty disks with increasing complexity}
\label{Model:models}

\begin{figure}
\centering
\includegraphics[trim = 0mm 90mm 95mm 0mm, clip, width=12cm]{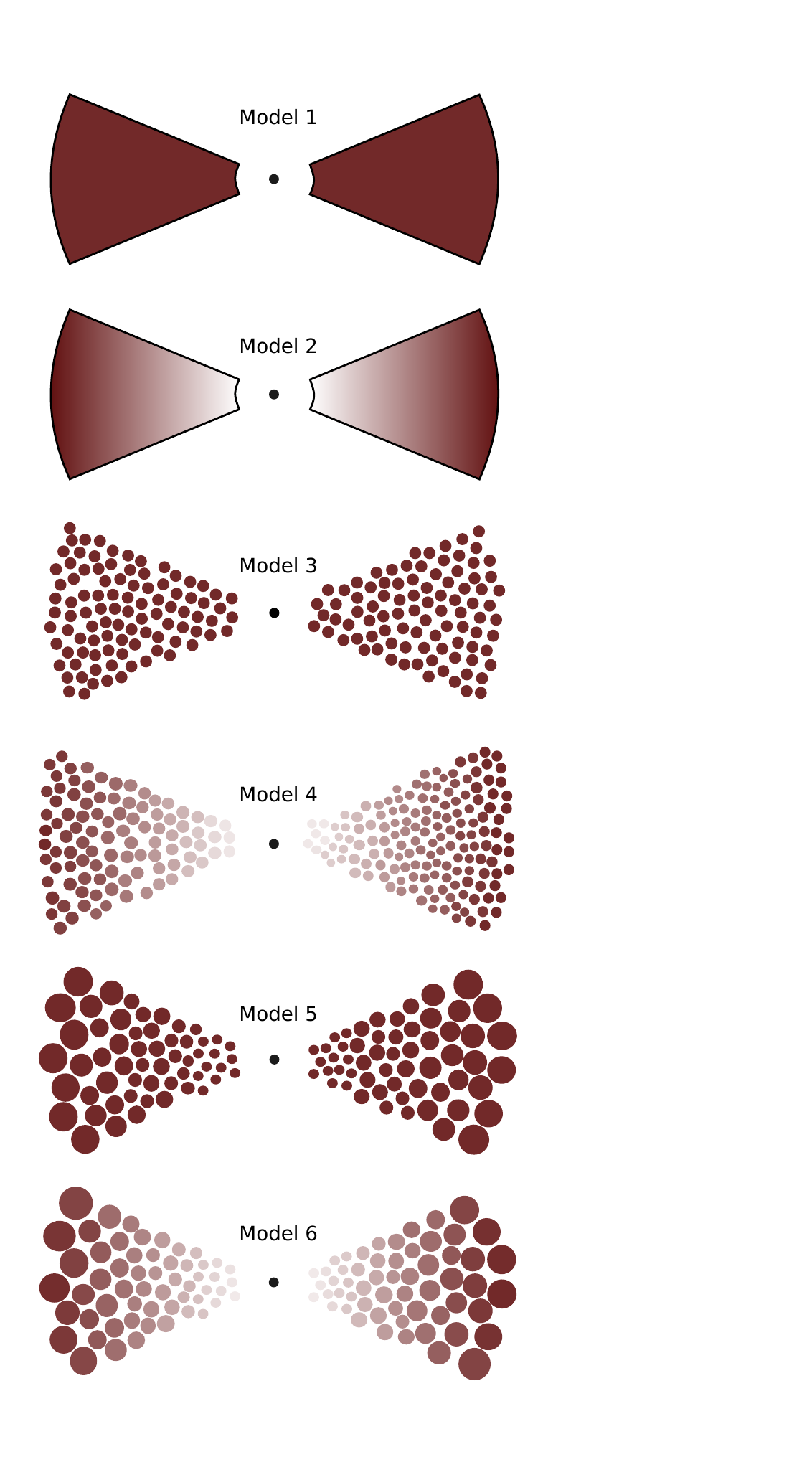}
  \caption{Unscaled geometrical and density variations for our models 
	    of dusty disks. The first two panels show a model 
	    with a uniform geometry and the last four panel
	    a set of fragmented disks. A uniform red filling 
	    indicates a constant optical depth, while a color 
	    gradient indicates a variation in the dust density 
	    with increasing radial distances from the central 
	    source. Details are given in the text.}
  \label{Fig:Geometries}%
\end{figure}

\begin{figure}
\centering
\includegraphics[trim = 10mm 5mm 5mm 5mm, clip, width=10cm]{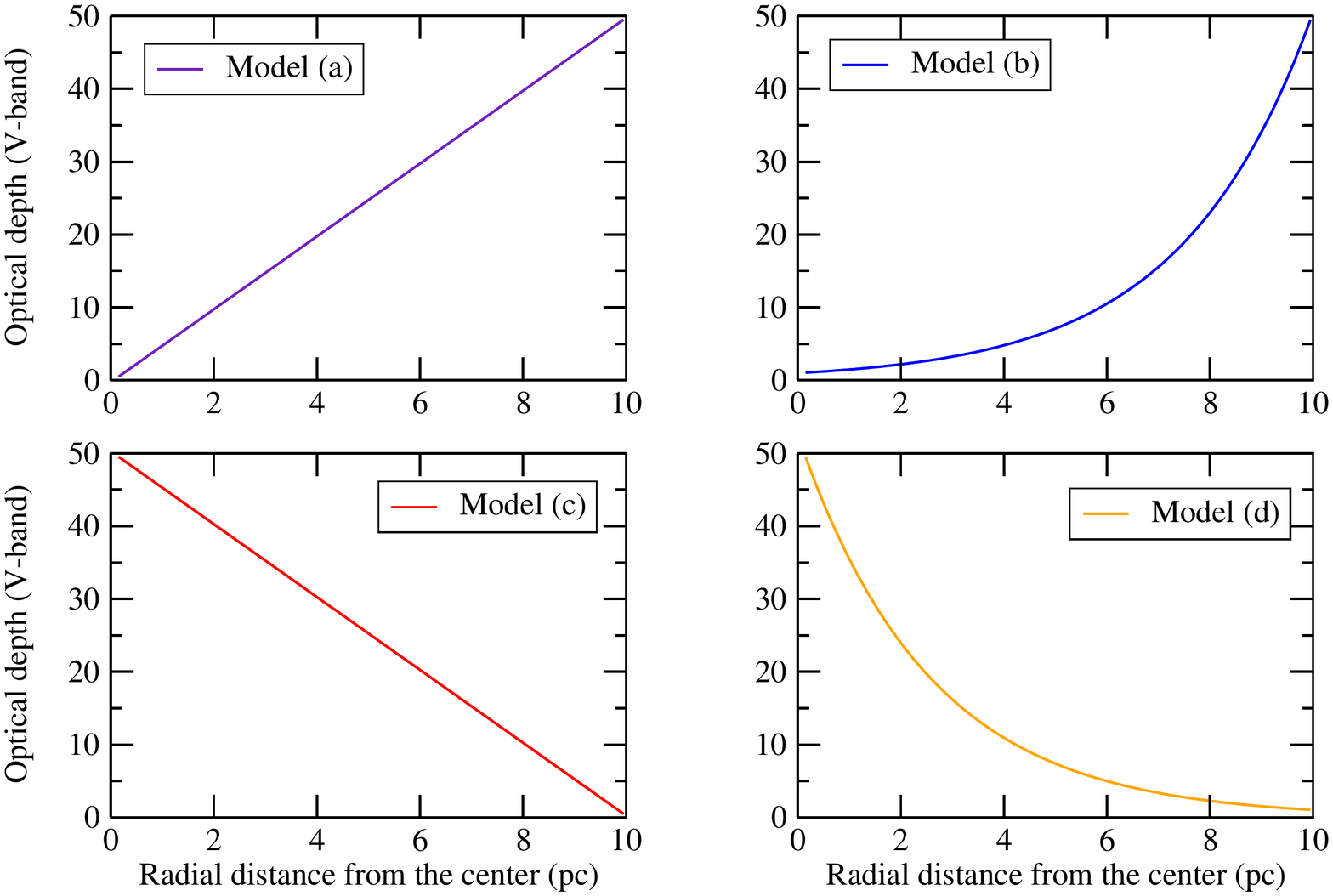}
  \caption{Theoretical dust distributions corresponding to 
	   heterogeneously-filled dust disks. Each model is 
	   color-coded and shows a different variation of 
	   the radial optical depth (in the V-band) with 
	   increasing distance from the central source. 
	   Details are given in the text.}
  \label{Fig:Optical_depth}%
\end{figure}

We show in Fig.~\ref{Fig:Geometries} the six different disk geometries we investigate in this paper. We consider first a 
plain flared disk that is uniformly filled with dust. Its radial optical depth in the V-band is larger than 100 to 
ensure a strong obscuration along the equatorial plane \citep{Antonucci1993}. This is the simplest configuration and 
it is one of the most used in numerical simulations. Model number 2 is also a plain flared disk but its dust density 
varies with radial distance from the central source. The third model is a clumpy flared disk with clumps of equal sizes 
and density. Each clump has an optical depth of 50 \citep{Siebenmorgen2015} and 7 to 10 clumps are obstructing 
the observer's line-of-sight along equatorial views, ensuring a total optical depth equal to or larger than 100. Its 
non-homogeneously filled counter-part is model 4, where the dust density varies with the radial distance from the center. 
Finally, model number 5 is a clumpy disk with clump sizes becoming bigger and bigger as we move away from the center. A 
clump located at the outer part of the structure is ten times larger than a clump located at the inner edge. In this model
the dust density is inversely proportional to the clump radius so that each cloud has a constant optical depth of 50 in 
the V-band. The last model is the same as model 5 but the dust optical depth is no longer constant. 

The different distributions of dust used in models 2, 4 and 6 are presented in Fig.~\ref{Fig:Optical_depth}. We consider 
four dust distributions, defined by their optical depth in the V-band plotted as a function of the radial distance from 
the central black hole. The size of the disks being the same for all models, the change in optical depths is only due to
variations in dust densities. Model (a) corresponds to a linear increase of the optical depth with increasing distances 
from the central source. Model (b) is for an exponential growth. Model (c) and (d) follow the inverse trend, with 
model (c) being a linear decrease of the dust optical depth with increasing distances form the center, and model (d) an 
exponential decrease. We therefore have fifteen different test cases: model 1, models 2(a -- d), model 3, models 4(a -- d), 
model 5 and models 6(a -- d).

We sample 2000 clumps for each clumpy model, equating to a disk filling factor of $\sim$~25\%. The impact of the number 
of clumps has been studied in \cite{Marin2015} and it was found that the lower the filling factor, the lesser the 
observed polarization and the higher the total flux. Using a filling factor of $\sim$~25\% allows for a flux decrease 
between the polar and equatorial inclinations of, at least, one order of magnitude. For all models, we fix the dust 
composition to the Milky Way mixture \citep{Wolf1999}. This is the standard dust composition observed in the solar 
neighborhood, but different mixtures can reside in different systems, such as postulated by \citet{Gaskell2004} 
and \citet{Li2007}. In the Milky Way, the dust is essentially composed of 62.5\% carbonaceous dust grains and 37.5\% 
astronomical silicate. Both ortho- and para-graphite are considered, with twice as much para-graphite as ortho-graphite 
\citep{Wolf1999}. The grain size distribution ranges from 0.005~$\mu$m to 0.250~$\mu$m in radii, with a distribution 
$n(a) \propto a^s$ and $s = -3.5$. Varying the dust grain radius distribution or the dust composition will result 
in a different albedo, hence different scattered/absorbed fluxes and polarization. However, in the following, we 
always compare models with the same dust mixture. The exact dust prescription is then of no importance since we are 
looking at model differences and not quantitative values. Finally, $i$ is the disk inclination, i.e. the angle 
between the observer line of sight and the symmetry axis of the disk. Pole-on view corresponds to $i=0^\circ$, 
edge-on view to $i=90^\circ$.

\subsection{Polarized radiative transfer}
\label{Model:STOKES}

We used the radiative transfer code {\sc stokes} to achieve our simulations. Developed by \citet{Goosmann2007} and 
upgraded by \citet{Marin2012,Marin2015}, {\sc stokes} is a Monte Carlo code that simulates the propagation of radiation 
in a user-defined three-dimensional environment. The code is extensively detailed in the three papers of the 
series and we only give here a brief summary of its important features. A list of applications of {\sc stokes}
can be found in \citet{Marin2014}.

{\sc Stokes} can model a large range of geometrical shapes that can be filled with electrons, dust grains, atoms 
and molecules. Emitting regions can be parametrized as sources with a given spectral shape. Photons, 
once emitted, travel along straight lines until a reprocessing event happens, depending on the optical depth 
of the media (calculated as $\tau$~=~$\displaystyle \int_{0}^{l} \sigma n(z) \, \mathrm{d}z$, $n(z)$ being the
number density of that material at $z$, and $\sigma$ the attenuation cross section). Photons can scatter, be absorbed 
and be re-emitted in all directions according to the Mie, Rayleigh, Thomson or Compton laws. Multiple scattering can 
occur and the four components of the Stokes vector are modified at each scattering-event. A spherical web of virtual
detectors registers each escaping photon, saving its polar and azimuthal coordinates together with the intensity, 
polarization (linear and circular) and time-lag. The polarization we measure is the scattering-induced, linear, 
continuum polarization degree ranging from 0 (unpolarized) to 100\% (fully polarized). The code can be parallelized 
and it takes a few days on a basic computer for a computation such as presented in this paper to sample 10$^{10}$ photons.

In the following, we ran our models for three different wavebands: near-infrared (mono-energetic photon emission
centered at 10~000~\AA), optical (5500~\AA), and ultraviolet (2000~\AA). Since the scattering phase function of dust 
is energy-dependent, we aim to find which is the best suited waveband for a polarimetric detection of the specific 
features of complex dusty disks. The polarimetric results we present are entirely due to scattering. We 
do not account for external dilution by the host galaxy, stellar starlight or thermal reprocessing by dust grains.
This might have an effect on the polarization signatures, particularly in the infrared domain where dilution by the 
host galaxy is much more important than in the ultraviolet band (see, e.g., \citealt{Bolzonella2000}), and it will 
be discussed in Sect.~\ref{Discussion}. We registered the flux, polarization and time-lag along 40 polar directions 
(from the symmetry axis of the model to the equatorial plane) equally spaced in cosine. We azimuthally integrated 
the signal to get better statistics. It was shown in \citet{Marin2015} that, even for a clumpy distribution, an 
azimuthally-averaged value is representative of the system provided it is not highly non-axisymmetric.

\section{Results}
\label{Results}

\subsection{Uniform disks}
\label{Results:Uniform}

\begin{figure*}
\centering
\includegraphics[trim = 5mm 5mm 5mm 5mm, clip, width=16cm]{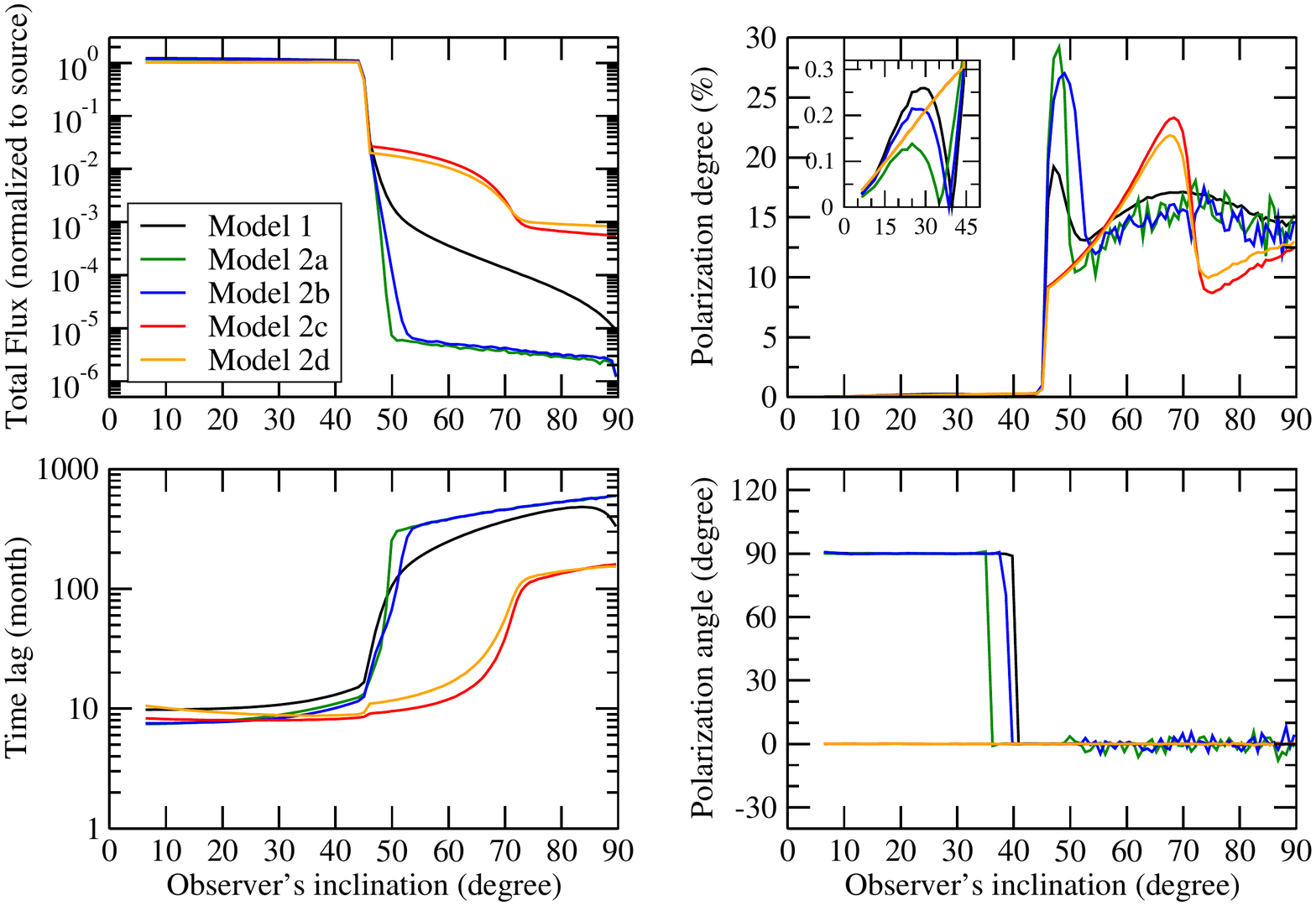}
  \caption{Near-infrared (10~000~\AA) total flux (normalized to 
	  the source emission), polarization degree, polarization 
	  angle and time-lag (normalized to the size of the disk)
	  as a function of the observer's inclination. Five plain 
	  flared disk models with different dust distributions are 
	  presented (see Sect.~\ref{Model:models} for details).}
  \label{Fig:Uniform_IR}%
\end{figure*}

\begin{figure*}
\centering
\includegraphics[trim = 5mm 5mm 5mm 5mm, clip, width=16cm]{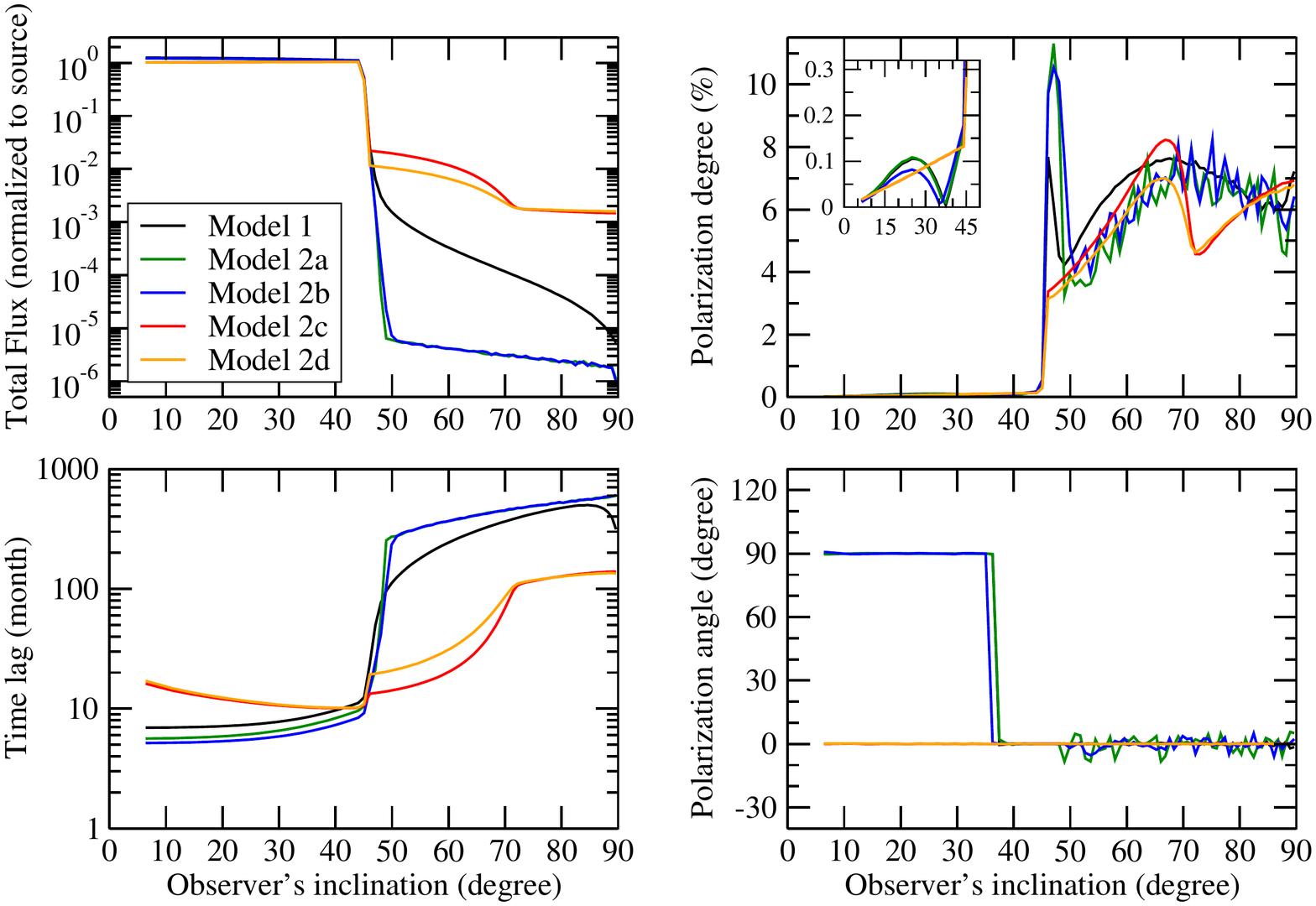}
  \caption{Same as Fig.~\ref{Fig:Uniform_IR} 
	   but in the optical band (5500~\AA).}
  \label{Fig:Uniform_OPT}%
\end{figure*}

\begin{figure*}
\centering
\includegraphics[trim = 5mm 5mm 5mm 5mm, clip, width=16cm]{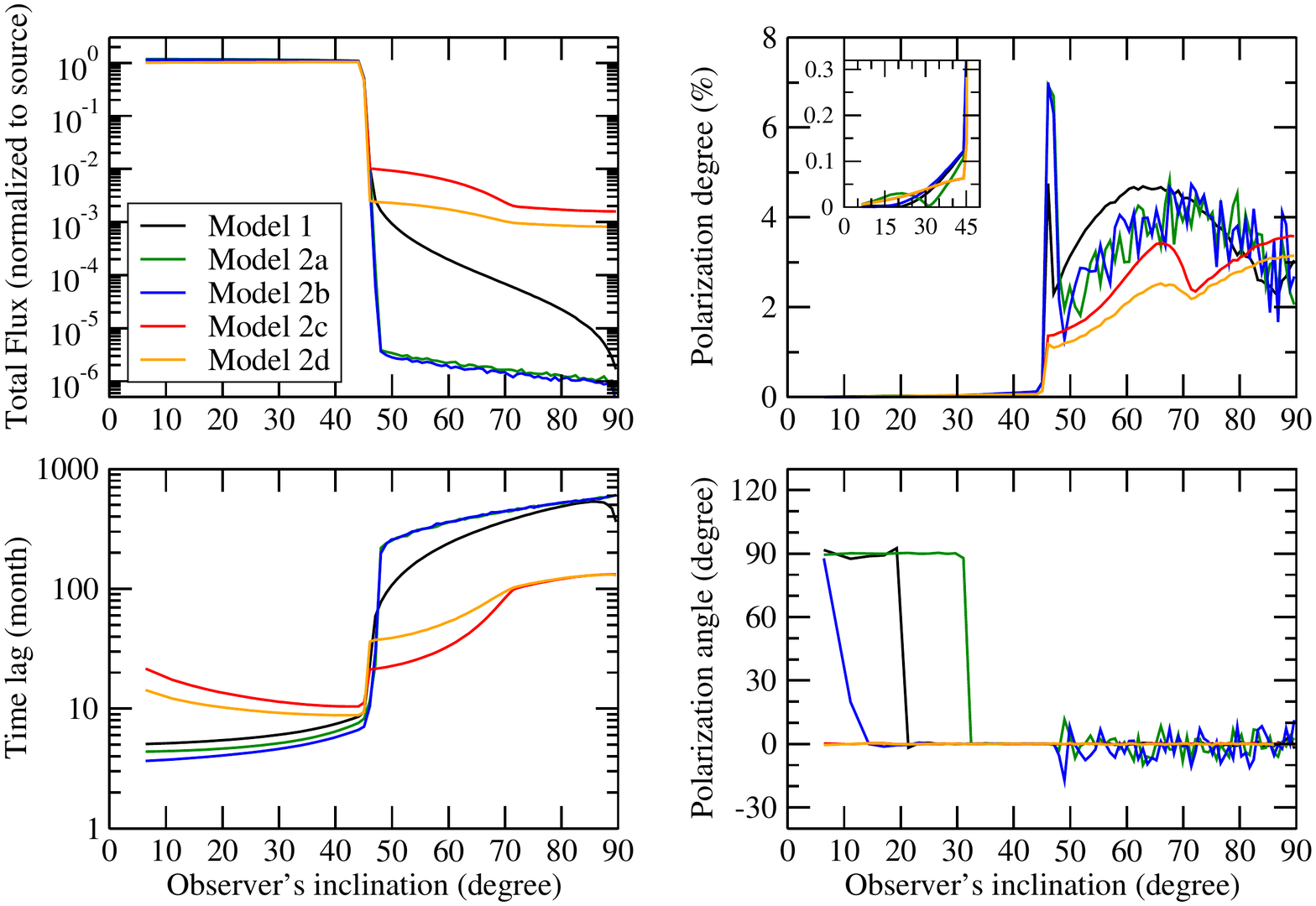}
  \caption{Same as Fig.~\ref{Fig:Uniform_IR} 
	   but in the ultraviolet band (2000~\AA).}
  \label{Fig:Uniform_UV}%
\end{figure*}

We present our first results in Fig.~\ref{Fig:Uniform_IR}, Fig.~\ref{Fig:Uniform_OPT} and Fig.~\ref{Fig:Uniform_UV}
for the near-infrared, optical and ultraviolet bands, respectively. They corresponds to models 1 (plain, uniformly-filled
dusty flared disk) and 2 (plain disks with radial dust stratification). We investigate, in model 2, all four dust 
prescriptions (a), (b), (c), and (d) according to Fig.~\ref{Fig:Optical_depth}.

\subsubsection{Radiation path}

\begin{figure}
\centering

\includegraphics[trim = 0mm 190mm 88mm 0mm, clip, width=10cm]{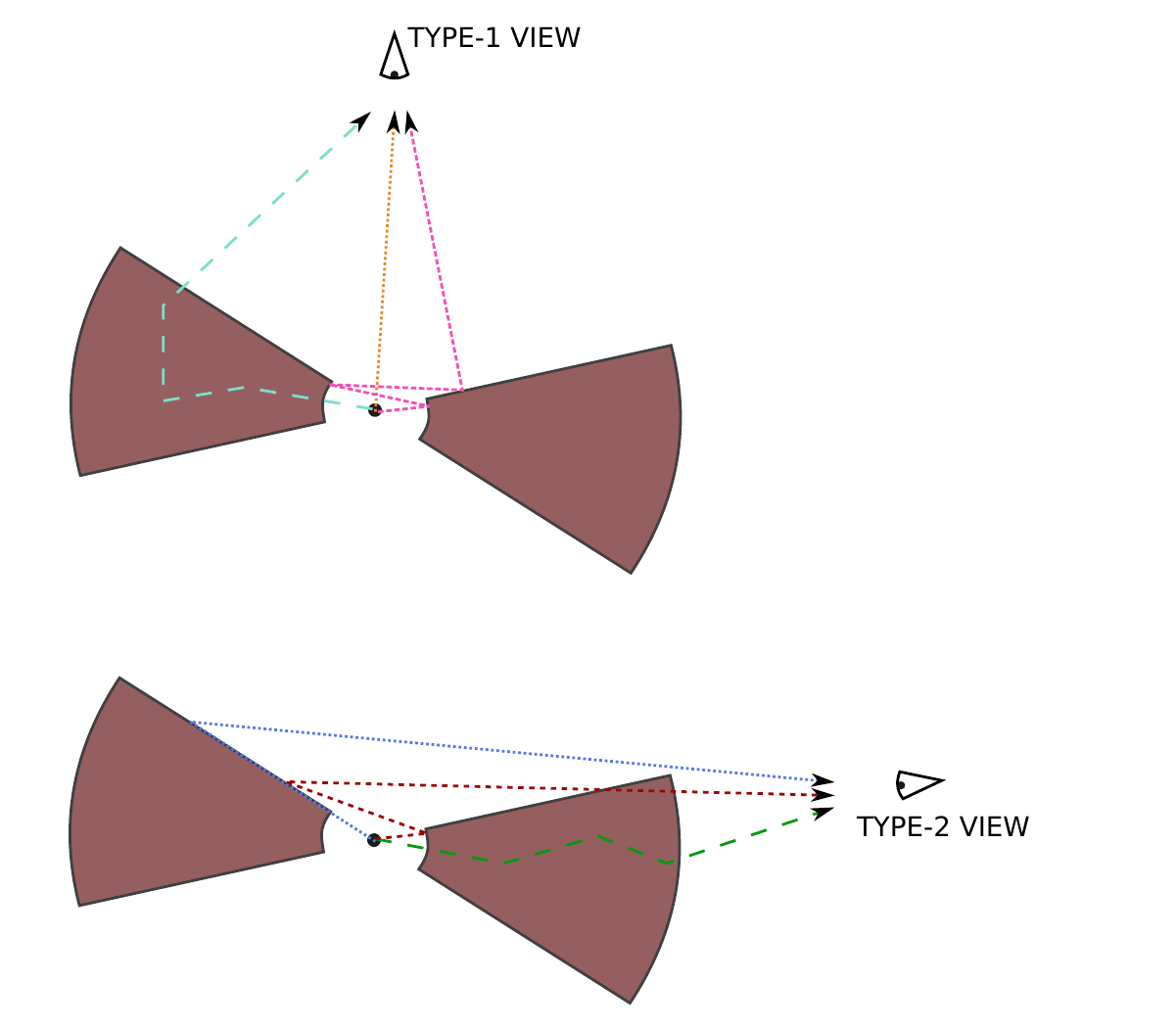}
  \caption{Illustration of different photon paths due to multiple 
	   scattering. See text for details. Top: pole-on (type-1)
	   view, bottom: edge-on (type-2) view.}
  \label{Fig:MultipleScatt}%
\end{figure}

Because of the large disk optical depth resulting in multiple scattering, the radiation path is not straightforward.
Photons may scatter more than twice before reaching the observer, especially if the disk is clumpy. We illustrate in 
Fig.~\ref{Fig:MultipleScatt} the different trajectories the photons can have. 

The top panel illustrates the case of an observer situated almost along the symmetry axis of the system, i.e. along 
the polar direction. In this case, there is a direct view of the central source and photons may travel freely towards
the observer (dotted orange line). This is the direct flux, which is unpolarized. If photons are emitted along 
the equatorial plane, they will encounter the disk funnel (dashed pink line). If not absorbed, they may scatter 
backward and be reprocessed onto the opposite funnel side until being absorbed or being able to escape. Due to the 
non-isotropic scattering phase function of Mie scattering, photon can also penetrate further into the disk (long-dashed 
cyan line). Radiation will scatter at different position before escaping at large distances from the central source. 
Depending on the position of the last scattering event, its polarization position angle can be either parallel or
perpendicular. For the remainder of this paper, we consider that a polarization position angle of 90$^\circ$ indicates 
a polarization angle parallel to the projected symmetry axis of the disk. This is the signature of equatorial scattering.
A polarization angle of 0$^\circ$ is thus perpendicular to the model axis and traces scattering event happening 
preferentially close to the pole of the model. Note that when a model is perfectly axisymmetric, the polarization 
angle can only take two values, 0$^\circ$ or 90$^\circ$. Otherwise, as in the case of a distribution of clumps, various 
polarization angles can be obtained, related to the statistics of the clumps. 

The bottom panel of Fig.~\ref{Fig:MultipleScatt} illustrates the case of an almost edge-on inclination. The central 
source is obscured by dust and no longer visible. Photons can escape via multiple scattering along the equatorial
plane if the dusty medium is not too optically thick or if there are gaps between the clouds (green long-dashed line).
The majority of radiation will preferentially scatter onto the disk surface opposite to the observer and then reach the 
detectors. This could happen with only one scattering event, where photon is emitted at an angle coincident to the 
half-opening angle of the disk (dotted blue line), or due to multiple scattering (dashed red line). When the inclination
of the system is close to 90$^\circ$, the obscuration is maximal and the amount of photons that have backscattered onto 
the disk surface opposite to the observer (blue and red lines) is minimal.

\subsubsection{Total flux}
Focusing first on the normalized photon flux (top-left panel of Figs.~\ref{Fig:Uniform_IR}, \ref{Fig:Uniform_OPT} and 
\ref{Fig:Uniform_UV}), we see an almost inclination-independent flux value until the observer's line-of-sight becomes 
obstructed by the dusty disk. The central source being unobscured, there is little contribution from the flared disk
that absorbs/scatters a small fraction of radiation. Nonetheless, models 2(c) and 2(d) present a slightly smaller 
(by a factor 1.3) flux at those inclinations. The optical depth being maximum at the inner disk radius (see 
Fig.~\ref{Fig:Optical_depth}), radiation is more heavily absorbed. Thus, there is less photon scatter toward the 
polar axis than for the other cases. Once the the dusty material intercepts the observer's line-of-sight, the flux 
drops by several orders of magnitude. For a plain disk where density decreases with radial distances from the center 
(models (c -- d)), the minimum flux is about three orders of magnitude lower than at face-on inclinations. Radiation 
that successfully crossed the first, dense layers of the inner disk region can freely propagate towards the disk outer 
radius. The number of photons is thus higher than in the case of a disk with a constant-density profile (model 1). 
Here, the flux drops by five orders of magnitude and the attenuation is almost linear with inclination. The attenuation 
is even stronger for the last two models (a -- b) when the dust density increases with radial distances from 
the source, the flux obscuration is very efficient. Between $i=45^\circ$ and $i=50^\circ$, the flux sharply drops
by five orders of magnitude and then stays constant at edge-on inclinations. In this specific case, photons have 
traveled far inside the dusty material before encountering the densest layers of dust. Radiation then becomes trapped 
inside the dust disk and naturally suffers from heavy absorption caused by multiple scattering. The emerging flux 
at edge-on inclinations is thus extremely weak as compared to the source. 

As the photon energy increases, models (c -- d) with decreasing radial optical depth from the center of the disk 
show a slightly higher edge-on normalized flux, while it is the opposite for the other cases. This complex behavior
is related to the albedo and the scattering and extinction cross sections of dust. The albedo is flat in the optical 
band but decreases in the ultraviolet and infrared regimes. On the other hand, the scattering cross-section decreases 
regularly with wavelength. An illustration is given in Fig.~3 in \citet{Goosmann2007}. The combination of the two 
factors allows ultraviolet radiation to cross a thicker layer of dust close to the inner radius of the disk: more 
ultraviolet photons can penetrate the disk, but they get more scattered or absorbed than optical or infrared photons.
However, in the case of a model with increasing dust density with radial distance from the source, the first layers 
of dust being optically then, the same ultraviolet photons will have traveled farther inside the dusty material before 
being absorbed or scattered. Scattered photons will more easily escape along the polar direction, where the remaining 
optical depth is smaller, rather than towards the equatorial plane where the optical depth is higher. This results in a 
decrease of the observed flux even at higher photon energies.

\subsubsection{Polarization properties}
The polarization properties of the five models (right panels of Figs.~\ref{Fig:Uniform_IR}, 
\ref{Fig:Uniform_OPT} and \ref{Fig:Uniform_UV}) strongly differ both in terms of polarization degree and polarization 
position angle. At polar inclinations, the degree of polarization is always small, less than a percent. This is due 
to the strong dilution effect of the unpolarized primary source that outshines the reprocessed component. The 
polarization degrees are marginally dependent on the half-opening angle of the disk \citep{Marin2012} but are very 
sensitive to its composition. There are however differences between models, even for low inclinations. As it can be 
seen in the small zoom-box inside the polarization degree panels, a uniformly-filled dusty disk produces the highest 
polarization degrees, while a disk with a linear increase of the dust density with radial distance from the center 
has the lowest continuum polarization. If the dust density decreases with radial distances (models (c) and (d)), the 
degree of polarization linearly increases with inclination until $i=45^\circ$. From 0$^\circ$ to 45$^\circ$, the other 
three models show a non-linear variation of their polarization properties, a behavior that is strongly impacted by 
the amount of reprocessed flux, hence by the model itself. This can be clearly seen when considering the intrinsic 
polarization position angle. When the dust density is constant or increases with radial distance from the center, 
the polarization angle is 90$^\circ$ whereas it is 0$^\circ$ otherwise. The first three models are dominated by 
scattering along the equatorial plane, producing a parallel polarization angle, until the line-of-sight of the observer 
is obscured by the dusty medium. Scattering then occurs along the polar directions, where photons reprocess on the 
flared disk surface opposite to the observer's position, see Fig.~\ref{Fig:MultipleScatt}. The polarization angle 
naturally rotates and the transition between the two polarization states causes the polarization degree to decrease 
around $i=45^\circ$ (the half-opening angle of the flared disk). When the disk optical depth decreases with radial 
distances from the source, the inner edge of the disk is similar to a thick wall of dust and radiation preferentially 
scatters along the polar direction if it is to escape from the disk, hence the perpendicular polarization angle. 

As soon as the source is no longer directly visible by the observer, the polarization angle is fixed for all models
(0$^\circ$) and the polarization degree rises due to Thomson laws. In the case of a uniformly-filled plain flared disk, 
the maximum infrared polarization degree is $\sim$~20\% at $i=50^\circ$, where polar scattering dominates. The polarization 
then decreases with increasing inclinations, as it becomes more and more difficult for radiation to escape the dust 
funnel, except by multiple scattering which will induce a natural depolarization effect. In the case of models 2(a) 
and 2(b), the maximum of polarization (27 -- 29\%) is reached for inclinations close to 45$^\circ$. The polarization 
degree then decreases for the same reasons as before, but the final edge-on value is lower than in the previous case. 
This is due, as explained above, to the confinement of radiation inside the flared disk and the subsequent heavy 
absorption. It also explains the lower statistics at edge-on inclinations. Finally, for the last two models in which 
the dust density decreases with radial distances from the center, the polarization degree peaks at inclinations close 
to 70$^\circ$. The maximum infrared polarization is as high as 23\%, and then decreases at equatorial orientations, with 
a polarization level similar to the uniformly-filled case. We thus see that the maximum polarization degree from a 
plain flared disk occurs for different inclinations and strongly depends on the radial distribution of dust. We see 
similar behaviors in terms of polarization angles and polarization degrees for all three wavebands considered here, 
with the polarization degrees decreasing at bluer wavelengths. We also note that the rotation of the polarization 
position angle of model 2(a) occurs at slightly different inclinations for different energies. Those two phenomena 
are due to the increase of the dust scattering cross section with decreasing wavelengths.

\subsubsection{Time lags}
The last panel of Figs.~\ref{Fig:Uniform_IR}, \ref{Fig:Uniform_OPT} and \ref{Fig:Uniform_UV} (bottom-left)
represents the averaged time-lag between actual photons reaching the observer and photons emitted by the central 
source that would not have suffered any scattering. Direct emission has a time-lag of zero. This quantity, 
with respect to the previous ones, is not directly observable as real cosmic photons are emitted at different times. 
The true averaged time lags must be reconstructed from observations by comparing polarized and primary light curves 
through theoretical transfer functions. The time-lags we present therein is to be used as a time indicator of the 
complexity of the radiation paths and the difficulty photons can have to escape from the dusty disk.

We see that, for pole-on inclinations, the time lag is rather small. The continuum is directly seen by the observer 
(zero time-lag) and polarized radiation mainly scatters on the inner disk funnel before reaching the observer, lagging 
the continuum by about 10 months. In the infrared band, the uniformly filled model shows the longest times lags but 
the increase with respect to the other models between 0$^\circ$ and 45$^\circ$ orientations is merely detectable. 
It is only at inclinations larger than 45$^\circ$ that the five models become truly unique in terms of temporal 
signatures. When the dust density increases with radius, the time-lag {abruptly increases when the inclination 
becomes slightly larger than} 45$^\circ$ and then becomes stable. Since there is a very small number of photons 
that can escape from the densest outer layers of the disk (see the flux plot), most of the radiation detected 
at those angles is due to the backscattering of photons onto the opposite disk edge. The photon travel length 
being almost similar at all edge-on inclinations, the time-lag stabilizes at 400 -- 600 months. A model with 
a constant-density dust prescription has a similar behavior but the time-lags are less important since radiation 
faces a constant, high optical depth and thus scatters many times before escaping but is not trapped in the 
outer regions, in contrast with the previous cases. The time-lag slightly drops at perfect equatorial inclinations 
since photons that have backscattered on the opposite disk funnel are not longer able to reach the observer 
and thus do not delay the averaged time response. Finally, for models with decreasing radial dust densities, 
time-lags remain quite constant before and after the transition angle, solely showing a small increase a 
inclinations larger than 70$^\circ$. This is also due to the amount of backscattered photons that outnumbers 
those that have traveled through the equatorial plane (i. e. with a smaller time-lag). 

We finally note that the time-lags differ in the near-infrared and the optical/ultraviolet bands, when 
energetic radiation can penetrate further in the dust material. For models with denser dust layers close 
to the source, IR photons mainly scatter onto the dust funnel while UV photons penetrate more easily in the
material, increasing their time-lag. The change in waveband affects the three other models in the opposite 
way: their time-lag at polar views is lower. This is due to the combination of albedo and scattering/absorbing 
cross sections variations, as explained above.

Summarizing what we have found for the plain flared disks, we have shown that different dust prescriptions 
result in specific inclination-dependent attenuations of the flux. If the dust layers become thicker 
with increasing distances from the central source, less photons are detected towards the equatorial regions.
It is not the case for the other configurations that allow more radiation to be detected at high inclinations.
The variations of the polarization degrees and angles are also very characteristic; the polarization is
maximum for models with radially-increasing dust densities. Those same models also provide the highest 
time-lags at equatorial views at all three wavebands. We note that the differences are less pronounced 
for type-1 (pole-on) inclinations, where the structure of the equatorial distribution of dust has a lesser 
impact on the averaged time lags. This prevents a very clear determination of the morphology of the disk 
from time reverberation studies that can be achieved only for low inclination objects, showing the importance 
of coupling timing and polarimetric studies to break degeneracies.

\subsection{Fragmented disks with constant-radius clouds}
\label{Results:Fragmented}

\begin{figure*}
\centering
\includegraphics[trim = 5mm 5mm 5mm 5mm, clip, width=16cm]{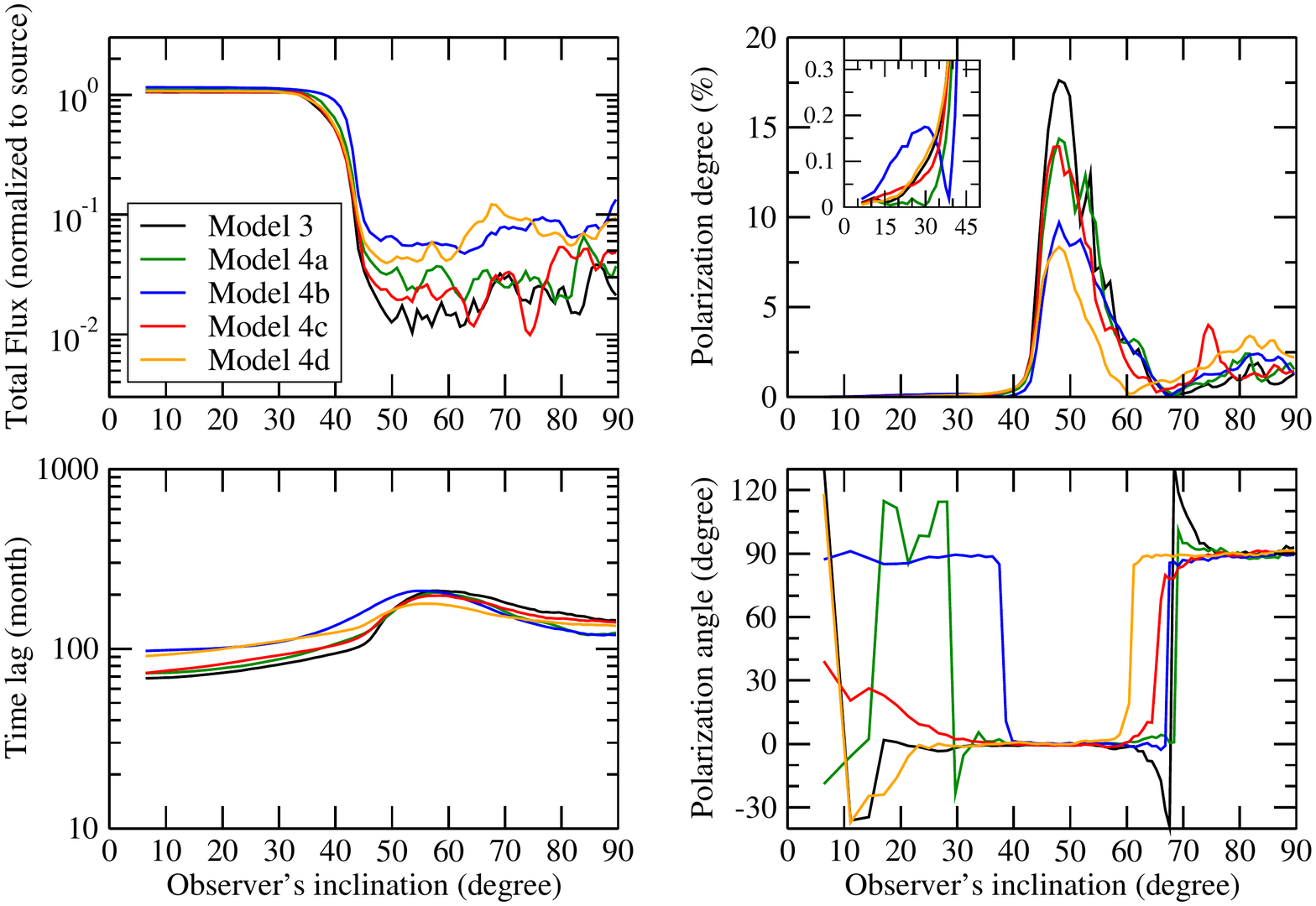}
  \caption{Near-infrared (10~000~\AA) total flux (normalized to 
	  the source emission), polarization degree, polarization 
	  angle and time-lag (normalized to the size of the disk)
	  as a function of the observer's inclination. Five clumpy 
	  flared disk models with clumps of constant radius but 
	  different dust distributions are presented (see 
	  Sect.~\ref{Model:models} for details).}
  \label{Fig:Fragmented_IR}%
\end{figure*}

\begin{figure*}
\centering
\includegraphics[trim = 5mm 5mm 5mm 5mm, clip, width=16cm]{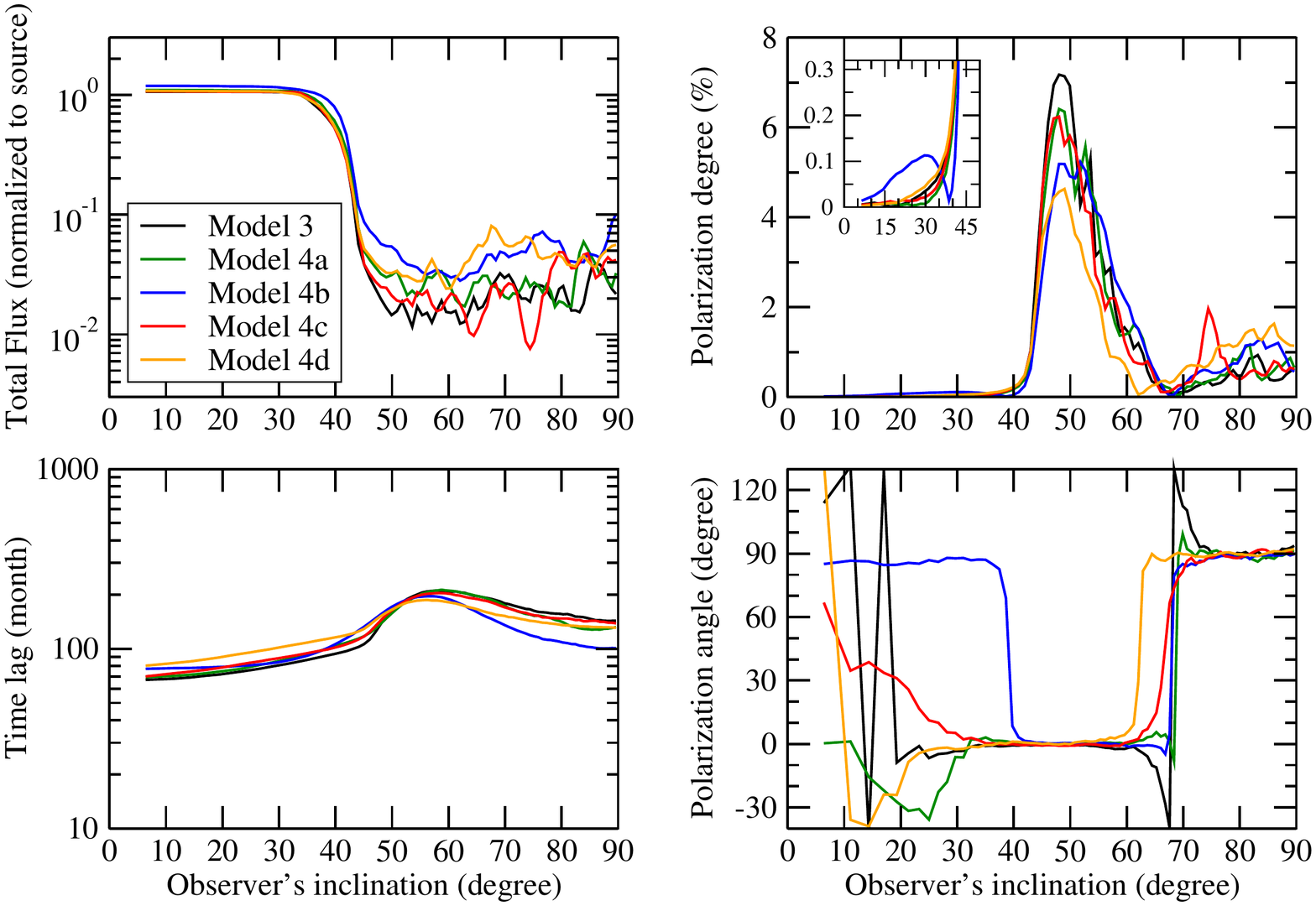}
  \caption{Same as Fig.~\ref{Fig:Fragmented_IR} 
	   but in the optical band (5500~\AA).}
  \label{Fig:Fragmented_OPT}%
\end{figure*}

\begin{figure*}
\centering
\includegraphics[trim = 5mm 5mm 5mm 5mm, clip, width=16cm]{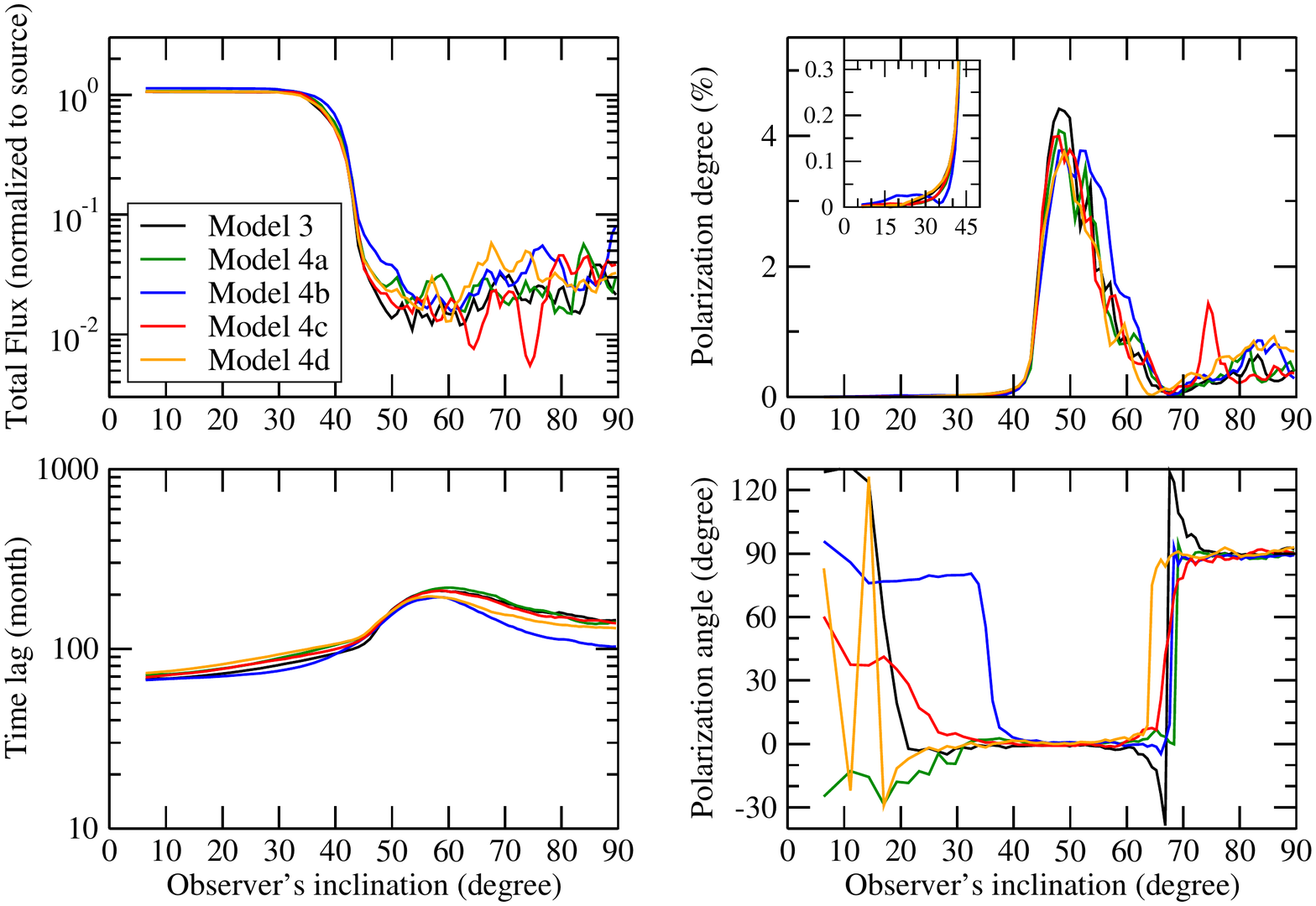}
  \caption{Same as Fig.~\ref{Fig:Fragmented_IR} 
	   but in the ultraviolet band (2000~\AA).}
  \label{Fig:Fragmented_UV}%
\end{figure*}

Our second series of models focuses on fragmented disks with clumps of equal sizes. We examine in 
Fig.~\ref{Fig:Fragmented_IR}, Fig.~\ref{Fig:Fragmented_OPT} and Fig.~\ref{Fig:Fragmented_UV} a model 
where the optical depth is the same for all clumps (fixed to 50 in the V-band, model 3) and a 
model where the dust opacity varies with radial distance from the central source (model 4). In the 
later case, we consider the same four dust prescriptions (a), (b), (c), and (d) as previously (see 
Fig.~\ref{Fig:Optical_depth}).

The normalized photon fluxes are maximum at polar inclinations as discussed in the previous section. 
However the flux attenuation due to obscuration is much less important at edge-on orientations. The 
clumpy structure of the disk is responsible for gaps between the cloudlets, allowing radiation to 
escape more easily thanks to multiple scattering. The interclump medium is pure vacuum here, but 
similar results are found for a clumpy two-phased medium with high-density clumps embedded in a 
low-density interclump dust \citep{Stalevski2012}. The flux attenuation is model-dependent. In the 
infrared, fluxes along the equator are smaller by about one order of magnitude than the polar value 
for models with exponential opacity distributions, and by almost two orders of magnitude in the 
linear case. The constant-density model (model 3) presents the lowest equatorial flux. We therefore 
see that the dust distribution has a profound impact on the quantity of transmitted flux at edge-on 
views. The photon fluxes are almost inclination-independent once the observer's line-of-sight has 
crossed the disk horizon, with a smooth transition between $i=35^\circ$ and $i=50^\circ$. The geometry 
of the system drives the shape of the curves and all five models appear similar (albeit their flux levels).
Moving towards shorter wavelengths has no impact on the shape of the curve but the derived photon
flux at edge-on inclinations is slightly different due to the combined effect of albedo and cross-section. 
All models appear to give equivalent fluxes along the equatorial plane, between one and two orders of
magnitude less than along the pole. 

The polarimetric results (right panels of Figs.~\ref{Fig:Fragmented_IR}, \ref{Fig:Fragmented_OPT} and 
\ref{Fig:Fragmented_UV}) are very different from the plain flared disk models, and are also characteristic 
of the model used. Whereas the infrared polarization degree along polar directions for the plain disk cases 
can reach $\sim$~0.25\%, here it is at best 0.17\% for the model with an exponentially-increasing dust 
opacity with radial distances from the source (model 4b). For this model, radiation mainly scatters in the 
equatorial plane until reaching the outer part of the disk. At this location, the dust density is maximum 
and the resulting polarization position angle is equal to 90$^\circ$. A clumpy disk with a linear increase 
of dust opacity with distance (model 4a) also presents a 90$^\circ$ polarization angle when observed edge-on, 
but its polarization degree is smaller. The seemingly chaotic variations of the polarization angle are due 
to the clumpiness of the medium and not to insufficient statistics. For the other models (model 3 and models 
4(c-d)), the level of linear polarization at polar inclinations is very low due to the inefficiency of the 
clumpy disk to scatter photons towards the observer. The resulting polarization angle is 0$^\circ$. This 
efficiency increases with inclination as the observer's line-of-sight becomes obstructed by clumps. The 
infrared polarization degree sharply rises up to 18\% in the case of constant-density clumps (model 3). 
For dust densities that decrease linearly (model 4c), the maximum polarization degree is 15\% and for the 
last two models that are much less capable of absorbing/scattering radiation, the maximum polarization is 
about 10\%. For all models, the polarization position angle is equal to 0$^\circ$ between 
40$^\circ$ -- 60$^\circ$ inclinations since polar scattering dominates. At larger inclinations, there is 
a rotation of the polarization angle for all models: the polarization angle becomes parallel while the 
polarization degree decreases. Scattering along the equatorial plane is now the easiest path to escape the 
clumpy dusty disk. This results in low infrared polarization degrees with 90$^\circ$ polarization position 
angles. This behavior was already revealed in \citet{Marin2015} and \citet{Marin2017} for constant-density 
clumpy models but we now show that their results also apply to more complex dust distributions. As for the 
plain disk discussed in Sect.~\ref{Results:Uniform}, we find a wavelength-dependent polarization variation, 
the maximum polarization degree of the models decreasing for shorter wavelength. This is due to smaller 
scattering cross section when the photon energy increases, which allows radiation to travel more easily 
through the dusty structure without interacting. 

The time-lags results (bottom-left panel of Figs.~\ref{Fig:Fragmented_IR}, \ref{Fig:Fragmented_OPT} and \ref{Fig:Fragmented_UV})
are also very different from their plain disk counterparts. Even at polar inclinations the time needed to escape 
from the dusty structure is large for photons that have scattered only once or twice. Since the medium is clumpy, 
radiation has traveled far inside the disk before being scattered towards the observer. The time-lag is thus about 
10 times larger than for a plain disk, where radiation mainly scatters inside the dust funnel. Note that this is not 
due to a higher albedo but rather to the existence of vacuum between the cloudlets that allows radiation to travel
without being absorbed. Model 4(b) presents the highest time-lags due to the exponential increase of optical depth 
with distance to the source. Only the outer parts of the disk are filled with high density, high opacity clumps. 
Photons thus travel a large distance before being scattered towards the observer. All other models show approximatively 
equal time-lags, the lowest values being obtained for the constant-density sphere model (model 3). When obscuration 
becomes efficient (at inclination larger than 45$^\circ$), the time-lag values increase by a factor $\sim$ 2 depending 
on the model. This increase correlates with the flux attenuation and the rise of polarization degree observed in other
panels. This is a clear indication for a maximum efficiency of obscuration and a large number of scattering events. 
Finally, when equatorial scattering dominates at edge-on inclinations, the time-lags decrease but stay larger than 
at polar angles. The dependence of time-lags with inclination is very similar at shorter wavelengths, with only a 
marginal quantitative decrease. 

We conclude that, for clumpy disks with constant radius spheres, the attenuation between a polar and an equatorial 
inclination is not very effective. The transition between the two extreme inclinations is smooth and operates around 
the half-opening angle of the dusty structure. If the half-opening angle is not the same for all observable sources, 
then it becomes impossible to distinguish between different dust stratifications. However polarimetry can reveal the 
distribution of dust with respect to the central source as the polarization degree reaches different maxima for 
different cases. The best waveband for observing the polarimetric signatures of clumpy tori is in the near-infrared 
where polarization levels larger than 10\% are expected for a narrow range of inclinations. Different dust structuring 
also imprint the polarization position angle with specific values at polar orientations. Finally, reverberation 
measurements hardly differentiate between the different models since they all give the same inclination-dependent 
curves. The time-lags they produce are too similar to be clearly distinguished.

\subsection{Fragmented disks with variable clump sizes}
\label{Results:FragVar}

\begin{figure*}
\centering
\includegraphics[trim = 5mm 5mm 5mm 5mm, clip, width=16cm]{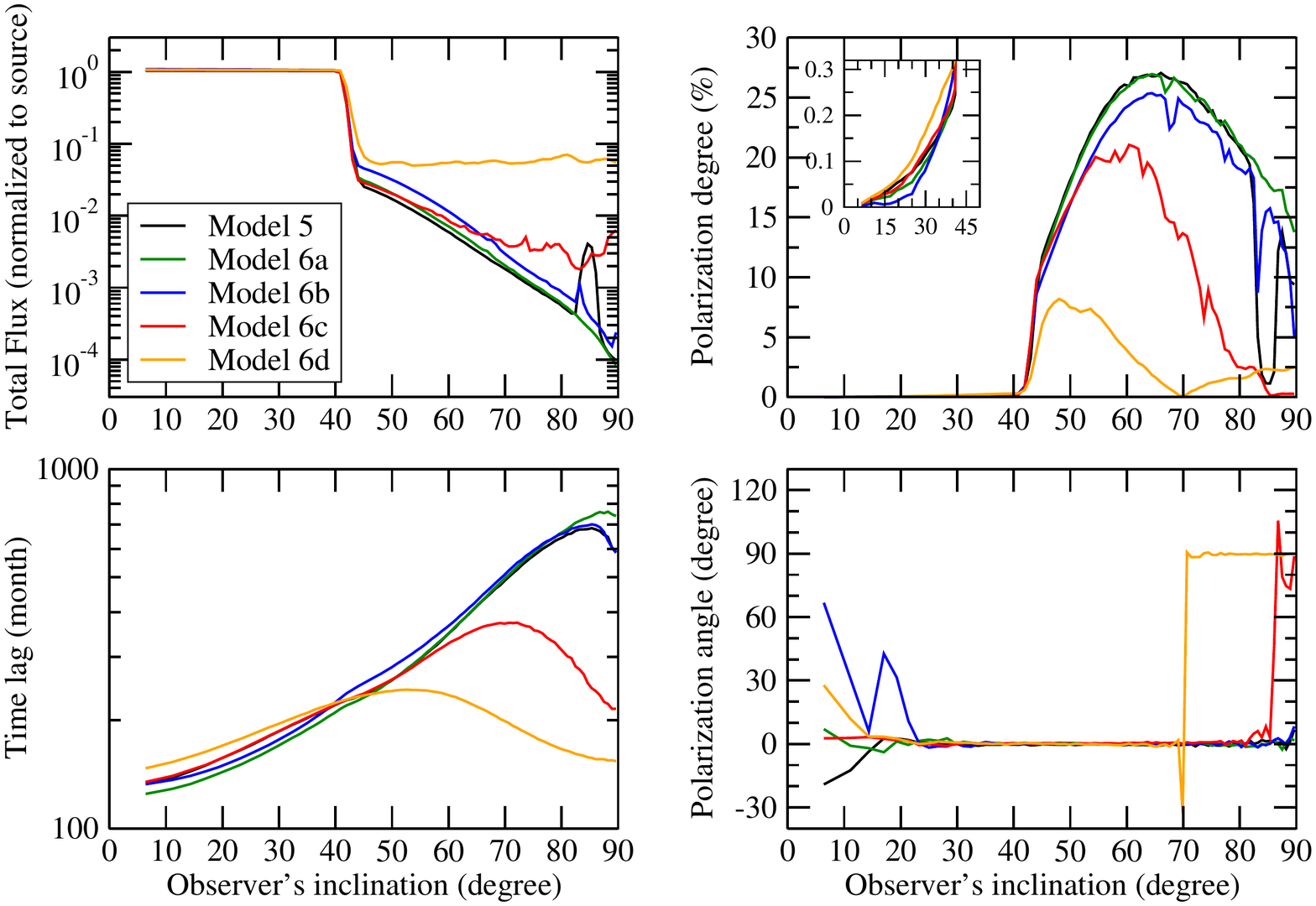}
  \caption{Near-infrared (10~000~\AA) total flux (normalized to 
	  the source emission), polarization degree, polarization 
	  angle and time-lag (normalized to the size of the disk)
	  as a function of the observer's inclination. Five clumpy 
	  flared disk models with clumps of increasing radius and 
	  different dust distributions are presented (see 
	  Sect.~\ref{Model:models} for details).}
  \label{Fig:FragVar_IR}%
\end{figure*}

\begin{figure*}
\centering
\includegraphics[trim = 5mm 5mm 5mm 5mm, clip, width=16cm]{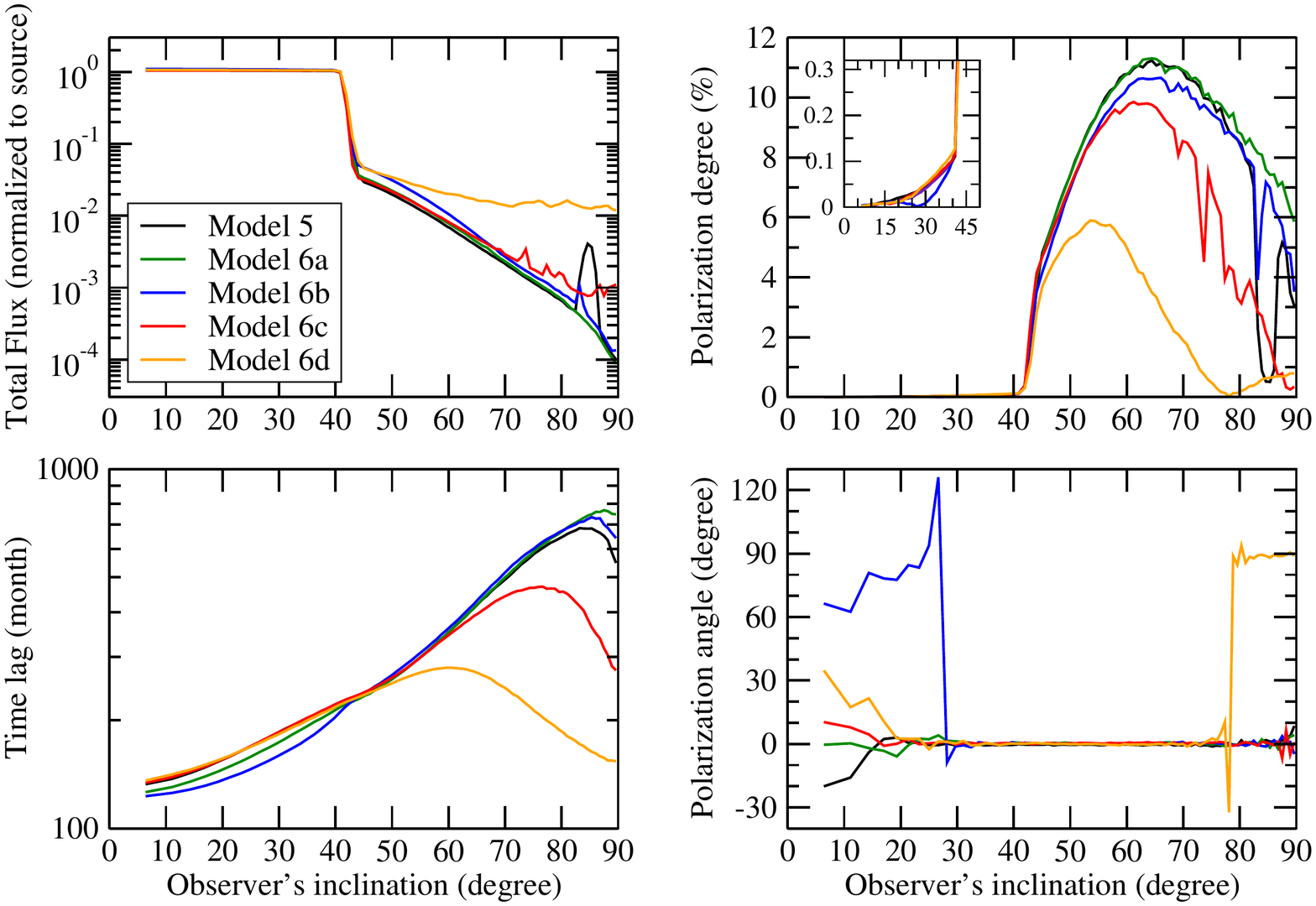}
  \caption{Same as Fig.~\ref{Fig:FragVar_IR} 
	   but in the optical band (5500~\AA).}
  \label{Fig:FragVar_OPT}%
\end{figure*}

\begin{figure*}
\centering
\includegraphics[trim = 5mm 5mm 5mm 5mm, clip, width=16cm]{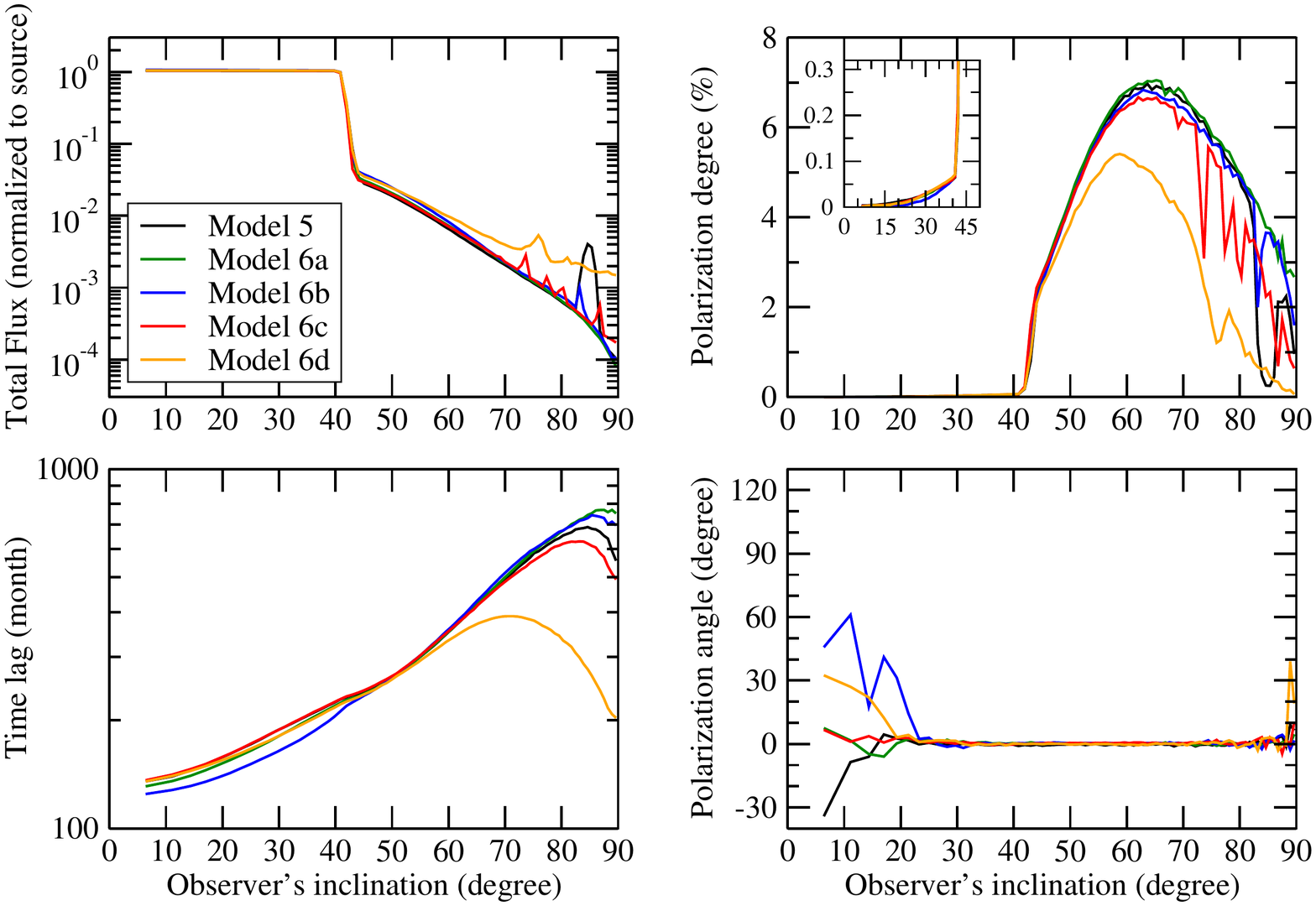}
  \caption{Same as Fig.~\ref{Fig:FragVar_IR} 
	   but in the ultraviolet band (2000~\AA).}
  \label{Fig:FragVar_UV}%
\end{figure*}

Our last set of models is presented in Figs.~\ref{Fig:FragVar_IR}, \ref{Fig:FragVar_OPT} and \ref{Fig:FragVar_UV}.
In this case, the dusty disk is clumpy and the clump radii increase with distances from the central source. We arbitrarily 
set the radius of clumps situated at the outer part of the structure to be ten times larger than clumps 
located at the inner edge. Similarly to Sect.~\ref{Results:Uniform} and Sect.~\ref{Results:Fragmented}, we examine a model 
where the dust opacity is constant for each clump (model 5) and four models (6(a) -- 6(d)) where the dust opacity varies
with radial distance from the source (see Fig.~\ref{Fig:Optical_depth}).

The normalized total flux appears to be constant at polar viewing angles since, as previously, it is dominated by direct flux 
from the central source. At low angles, the fluxes are inclination- and model-independent. Differences only appear when the 
observer's line-of-sight crosses the dusty disk. In the case where the opacity exponentially decreases with distance, 
the disk does not efficiently block radiation at equatorial viewing angles. The flux has decreased by only one order of 
magnitude. The gaps between the clouds and the presence of low opacity, large dust cloudlets at the outer disk radius do not 
allow for a large obscuration. If the dust opacity decreases linearly, obscuration is slightly more efficient since the cloud 
opacity is still higher than 10 (V-band) at 8~pc from the source (while it is about 2 -- 3 in model 6(d)). If the disk is 
composed of low opacity inner (small) clouds and high opacity outer (large) clouds, radiation is more heavily absorbed. Fluxes
are about four orders of magnitude lower at extreme equatorial inclinations for the three remaining models (models 5, 6a 
and 6b). The flux decrease with inclination is exponential for these three cases and the transition between unobscured and 
obscured sources is sharp at around the disk half-opening angle. The fluctuations we see in the flux curves are due to the 
distribution of clouds inside the model. For different realizations of clumpy disk models some features might appear if a given 
equatorial line-of-sight is not obscured by several optically thick clouds. This occurs, for example, in Fig.~\ref{Fig:FragVar_IR}, 
models 5 and 6(c), at inclinations larger than 70$^\circ$. The appearance of peaks and dips in the photometric curves are 
wavelength-dependent. As the photon energy decreases, radiation is less absorbed by dust grains and, if there is not enough 
dusty material to block radiation along a given direction, photon will escape more easily. This causes the low amplitude peaks 
of flux in the photometric curves observed in e.g. 6d which are visible in the ultraviolet but not in the infrared. 

The polarimetric plots (right panels of Figs.~\ref{Fig:FragVar_IR}, \ref{Fig:FragVar_OPT} and \ref{Fig:FragVar_UV}),
show a sharp rise of polarization with increasing inclinations. At polar angles, the polarization degree exponentially 
rises up to 0.3\%. The associated polarization position angle is 0$^\circ$ for all models in the near-infrared.
This means that photons have not scattered much along the equatorial plane before escaping the system, but have instead
mainly scattered on the large dust clumps located high above the equatorial plane, resulting in the characteristic 
perpendicular polarization angle. When the observer's line-of-sight becomes obstructed by dust, the polarization degree 
sharply rises since polar scattering clearly dominates over unpolarized emission from the (now) obscured source. Models
with decreasing optical depth with radial distances from the center (6c,6d) have the lowest polarization degrees, since photons
easily escape from the system, and therefore do not carry large scattering-induced polarization. This effect is even more 
pronounced at equatorial inclinations when scattering along the equatorial plane dominates over polar scattering, resulting 
in a rotation of the polarization position angle for the two aforementioned models. The polarization degree thus shows dips, 
with minima associated to the rotation of the polarization angle. The case is slightly different for the other three models 
(5, 6a, 6b). Their polarization degree is as high as 25\% and it is always associated with a perpendicular polarization position 
angle. For increasing inclinations the polarization degree regularly decreases for the same reason as above but since 
obscuration by large, optically thick, outer clumps is more efficient, equatorial scattering does not dominate and the 
polarization position angle remains equal to 0$^\circ$. At shorter wavelengths, the results are similar but the polarization 
degrees are lower (down to 7\% for the most polarized models in the ultraviolet band) as obscuration and depolarization 
through multiple scattering dominate over transmission. 

The time-lags (bottom-left panel of Figs.~\ref{Fig:FragVar_IR}, \ref{Fig:FragVar_OPT} and \ref{Fig:FragVar_UV}) are also
quite different for different dust opacity prescriptions. The time-lags are similar for all models at polar 
inclinations, since they mainly result from scattering onto the large dust clouds at the outer radii of the fragmented disk.
However, once the inclination is larger than 45$^\circ$, the five models show different behaviors. Model 6(d) (density 
decreasing exponentially with distance) shows the lowest time-lags, with an infrared maximum around $i=50^\circ$. The 
time-lag decreases at equatorial inclinations as photons preferentially escape by scattering close to the equatorial plane, 
carrying a smaller time delay than photons that have backscattered onto the disk funnel and then inside the cloud distribution. 
The same conclusions apply to a model with a linear decrease of density. It follows that a clumpy disk with increasing cloud 
radii and decreasing optical depths would be, contrary to naive expectations, poor target for reverberation measurements. 
Better targets are models 5, 6(a) and 6(b). The time-lags increase with inclination, by factors up to up to 5. The presence 
of optically thick, large dust clouds in the outer borders of the system greatly facilitate time-lag observations since 
photons have to travel through the whole disk before being efficiently blocked and/or scattered by dust. This statement 
is even strengthened at shorter wavelengths where the time-lags are larger due to the ability of optical and ultraviolet 
photons to penetrate deeper in the dusty medium.

In summary, our last set of models has shown that clumpy disks with clumps of variable sizes and optical thickness have 
very distinctive features that are quite different for a variety of dust prescriptions. The importance of large and dense 
clumps at the outer edges of the disk is crucial for obtaining an efficient obscuration. They also enable large polarization
degrees and time-lags. If the disk becomes more transparent with radial distances from the source, obscuration is less 
efficient and we observe a rotation of the polarization angle with increasing inclinations.

\section{Discussion and observational prospects}
\label{Discussion}
We have shown that varying the morphology and/or the distribution of dust density in equatorial dusty structures has a 
profound impact on the observational properties (total flux, polarization degree and angles, time-lags) we have considered. 
The flux detected at equatorial inclinations is very sensitive to the structure of the disk: uniform or fragmented media 
result in flux levels that can be orders of magnitude different from those observed at polar viewing angles. 
Clumpy structures (either with constant or variable cloudlet radii) tend to be less efficient in blocking radiation 
along the equatorial plane unless the outer clouds are substantially bigger than the inner clouds. The resulting 
polarization is also distinctively different between the three morphologies and we found it to be much more sensitive 
to the radial distribution of dust than the photometric marker. The polarization degree is inclination-dependent and 
varies together with rotations of the polarization position angle that can trace the dust stratification. 
Finally, the averaged time-lags resulting from multiple scattering in different models is also a very distinctive 
signature that can probe the morphology of the scattering medium. Those two observational properties, polarization and 
time-lags, are the bases of the polarized reverberation mapping technique introduced by \citet{Gaskell2012}.

\subsection{Polarized reverberation mapping}
\label{Discussion:polarized_rev_map}
By comparing the amplitude of variability of the polarized flux to the amplitude of variability of the total flux 
in a blue passband with effective wavelength of 3600~\AA, \citet{Gaskell2012} found that the polarized flux of the 
Seyfert-1 NGC~4151 follows the total flux with a lag of 8 $\pm$ 3~days. This, together with the orientation of the
position angle of the polarized flux, enabled to constrain the effective size of the unresolved equatorial scattering 
region of the AGN, most probably composed of electrons. This is the first and unique example so far illustrating the 
importance of polarized reverberation mapping in resolving the innermost structure of AGN. Achieving those observations 
is time consuming and required about seven years of monitoring the total and polarized fluxes of NGC~4151. In contrast 
with the work of \citet{Gaskell2012}, who probed the inner broad line region, we have shown here that dust scattering 
results in longer timescales, of the order of several months for type-1 (polar) inclinations. It is therefore very 
important to select the optimal wavebands before starting such long observational campaigns.

Our results indicate that the near-infrared band seems to be the best to search for high degrees of polarization, 
since photons are not energetic enough to travel far into the dusty medium; instead, they are reflected (the albedo is 
large) and do not penetrate much the disk, leading to high polarization degrees. This applies to either polar, intermediate 
or equatorial inclinations. However, the reader must bear in mind that thermal re-emission from dust and starlight 
dilution (from the host galaxy in the case of AGN) may also be strong enough to wipe out the polarization signature 
\citep{Miller1983}. Hence, a more secure waveband to search for the polarization signal from dusty equatorial structures 
is ultraviolet, where starlight emission decreases in spiral and elliptical galaxies \citep{Bolzonella2000}. 

The reverberation mapping technique requires to observe both the continuum and the reprocessed spectra; the central 
source must be directly visible and the line of sight must not cross the obscuring dusty torus. This limits us to 
polar viewing angles, for which the time-lags are the shortest. In the AGN context, this is beneficial since the 
averaged time delays between variations of the polarized flux and of the total flux can be of the order of months 
(see Fig.~\ref{Fig:Uniform_IR} for example). We compared our results to the ones obtained by \citet{Almeyda2017},
who investigated the mid-infrared reverberation response of a clumpy torus, and found them to be in agreement with 
ours, even if the authors considered lower energies. Depending on the isotropic or anisotropic distribution of 
clumps in their torus models, \citet{Almeyda2017} found an averaged optical-to-infrared time-lag of 80 -- 150~days 
(i.e. 2.6 -- 4.9 months) at 3.6~$\mu$m. In comparison, our clumpy model with constant cloudlet radius and optical 
depth gives a time-lag of 3.9 months (cross-correlated between the total and polarized fluxes at 1~$\mu$m). 
Shorter delays, can be obtained in the ultraviolet band (a conclusion also shared by \citealt{Almeyda2017}).

We therefore conclude that polarized reverberation mapping is quite possibly better suited for ultraviolet/blue 
observations. The time-lags between the unpolarized continuum and the polarized reprocessed spectrum will be shortened, 
while the observed polarization degree will be less diluted by starlight contribution. The combination of the 
two techniques, leading to $polarized$ time reverberation studies, will strongly help to break degeneracies. Our 
simulations suggest that much more sophisticated models are less easy to distinguish/constraint, but the combination 
of observables can still serve to rule out certain scenarios. The main difficulty here is that for low inclinations, 
the polarized fraction is small whereas for large inclinations, the unabsorbed, direct flux is severely reduced. 
\cite{Gaskell2012} have nevertheless clearly established the feasibility of such an observational study for the case 
of NGC~4151, but one has to remember that it took them about seven years to complete the observations. Multi-wavelength 
analysis such as radio observations to probe the kinematics and composition of the circumstellar/circumnuclear region 
or hard X-rays, which are much less absorbed/scattered than the optical, might be useful to narrow down the uncertainties. 
A detailed treatment of the problem is deferred to a forthcoming paper where we intend to construct artificial light 
curves in both total and polarized light to measure realistic time lags, including both dilution by the host, interstellar 
polarization and sophisticated transfer functions.

\subsection{Towards more sophisticated models}
\label{Discussion:sophistication}
We have shown that two dusty structures with the same geometry but different dust grain distributions (e.g, a linear increase 
and an exponential increase of opacity with radial distance from the central source) are distinctively different in terms of 
polarization and timing properties, see Fig.~\ref{Fig:FragVar_UV} for example. However, those differences are less likely 
to be detectable in terms of photometry, since almost all models give the same inclination-dependent trend in flux attenuation. 
To detect a radial stratification of dust, time-resolved polarimetry appears to be the solution but it is unclear that this 
conclusion would also applies for vertical stratifications and for chemical variations of the dust grains with distance from 
the source. It has been known for long that the dust composition and temperature of stellar nebulae varies with distance 
from its young star \citep{Bally2006}; the same applies to galaxies, since the galactocentric dust mineralogy varies with 
distance and metallicity \citep{Giannetti2017}. We thus naturally expect circumstellar and circumnuclear dusty disks to be 
also more complex than considered here.

\begin{figure*}
\centering
\includegraphics[trim = 10mm 5mm 5mm 20mm, clip, width=16cm]{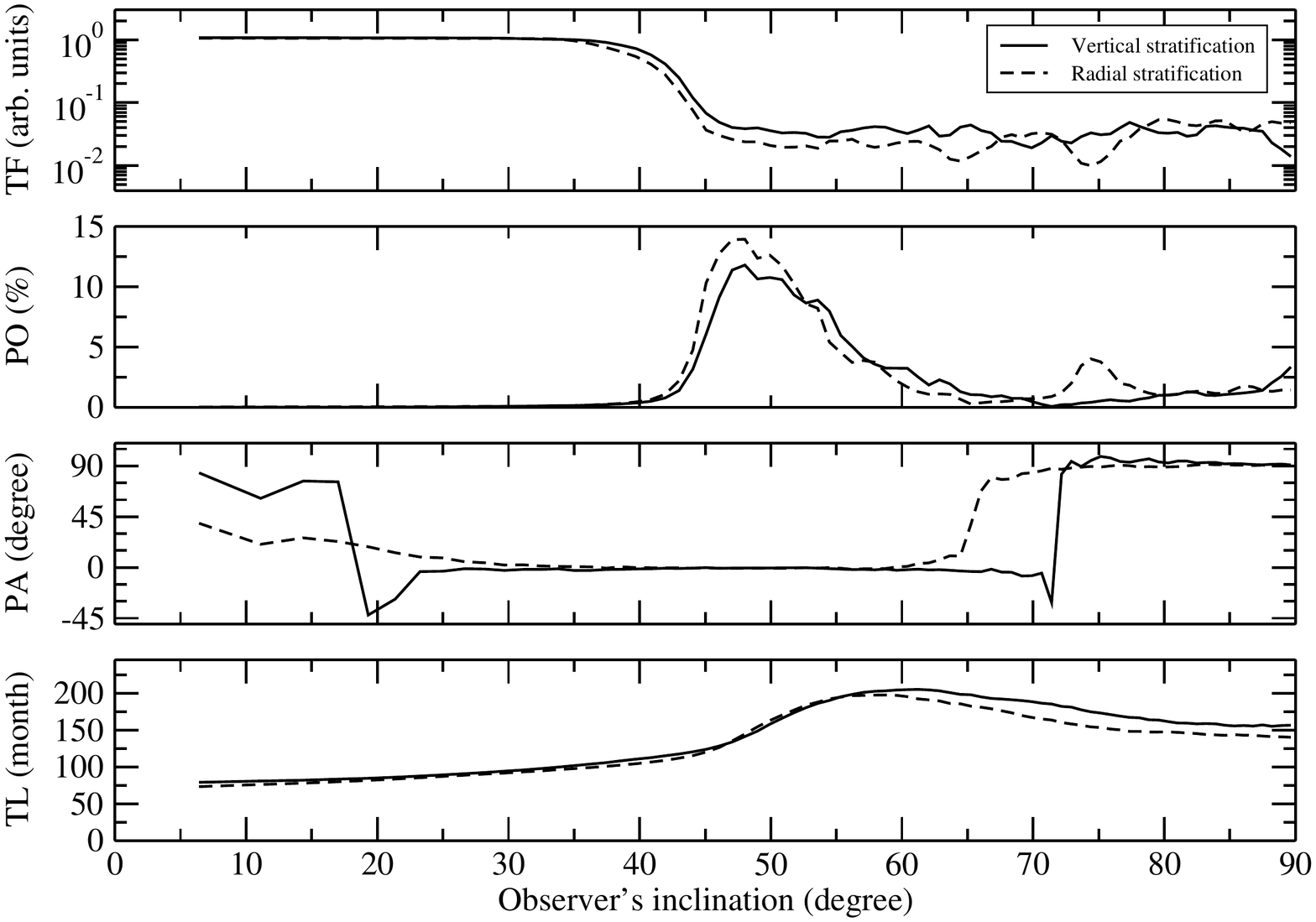}
  \caption{Comparison between a radial (solid dashed line) and a 
	  vertical (solid plain line) dust stratification in a 
	  clumpy equatorial flared disk with spheres of constant radius.
	  The variation of the dust density is linear and decreases
	  from the center of the model (first case) and from the 
	  equatorial plane (second case). The top panel shows the 
	  normalized photon flux at 1~$\mu$m, the second panel 
	  is the polarization degree, the third panel is the 
	  polarization position angle, and the last panel is the 
	  time-lag as a function of the system inclination.}
  \label{Fig:VerticalVar}%
\end{figure*}

It is beyond the scope of this paper to consider all possible dust distributions with distance, altitude, temperature 
or metallicity. We have already proven that uniformly filled equatorial dusty structures give different polarimetric and 
timing outputs than structures with radial dust opacity gradients. However, in order to illustrate the complexity of the 
parameter space for filling the disk with dust, we show in Fig.~\ref{Fig:VerticalVar} one realization of our numerical 
code for a clumpy disk with constant cloud radii in the case of a linear decrease of dust density with vertical distance 
from the equatorial plane. The simulation was achieved at 1~$\mu$m and results are shown in Fig. \ref{Fig:VerticalVar} 
using a plain solid black line. For comparison, we also show a model with the same parameters, excepts that the gradient 
is now radial instead of being vertical; the dust opacity linearly decreases with radial distance from the central 
source (black dashed line). As can be seen, the two configurations (vertical and radial variations of opacity) are 
almost indistinguishable in terms of photometry. The signal is the same at polar viewing angles and the obscured flux at 
equatorial orientations is equivalent. Differences can only be attributed to the random positions of the clouds. The unique 
true difference between the two models is the inclination angle at which the transition between the ``unobscured'' and 
``obscured'' state occurs. The vertical opacity stratification configuration is similar to a configuration where an optically 
thin atmosphere lays on top of denser equatorial layers. Photons grazing the disk surface are less likely to be absorbed. 
It is only at a larger inclination, where the medium finally becomes optically thick, that the disk can efficiently block 
radiation. In terms of polarization we see that the inclination-dependent degree of polarization follows the same trend, 
but the maximum amount of polarization is only 11\% instead of 14\% at intermediate inclinations. The small bump of 
polarization observed at an inclination of $\sim$~75$^\circ$ is due to the random distribution of the clumps position in 
the disk. The polarization position angle is the only marker showing a distinctive difference between the two configurations. 
The polarization angle is $90^\circ$ for $i=0^\circ$, indicating that scattering mainly happens along the equatorial 
plane. In comparison, for a radial opacity stratification, the polarization angle is close to 45$^\circ$ since both equatorial 
and polar scattering are contributing. Here, the vertical stratification of dust density favors scattering along the equatorial
plane where the disk is denser. For larger inclinations, the polarization angle rotates to 0$^\circ$, as in the radial case, 
since backscattering on the disk funnel becomes important at those intermediate inclinations. It is only at large inclinations,
when the continuum is obscured and when backscattering is no longer dominant with respect to equatorial scattering, that the 
polarization angle switches to 90$^\circ$. Similarly to the photometric results, the angle at which this transition happens 
is different from the radial stratification scenario. Finally, the inclination-dependent averaged time-lag is almost 
indistinguishable between the two cases since the geometry of the medium plays a more important role than its composition 
in time domain studies.

\section{Conclusions}
\label{Conclusions}
In this paper, we have shown that the morphology or the distribution of dust density of an equatorial dusty structure has a 
profound impact on the polarized signal and on the expected (averaged) time-lags. A plain uniform structure may have a similar 
behavior to that of a complex clumpy one in terms of flux attenuation but the polarization degree and angle would be 
significantly different. If the polarization signal from the equatorial dusty disk can be isolated from the central source or 
from additional components (which is more problematic in the case of AGN), it is possible, in principle, to distinguish between 
different models. In particular, the dust distribution inside a disk has a very clear inclination-dependent polarization signature. 
Since the inclination is not easy to determine from observations, coupling timing and polarimetric informations is crucial for 
determining the physical size of the unresolved region, together with its composition.

We also found that the ultraviolet/blue band is quite probably better suited for polarized time-reverberation studies. The 
polarization level due to dust scattering is not expected to be as high as in the near-infrared but the ultraviolet waveband 
should be less contaminated by external starlight and thermal re-emission. In addition, the time delays are expected to be 
slightly shorter in this waveband, allowing for less time-consuming observational campaigns. We intend to continue this study
by applying cross-correlation methods to light curves obtained by our Monte Carlo method. We aim at reproducing the observed 
time-delays of NGC~4151 first, estimate the inclination of the AGN by testing all the possible orientations thanks to our 
numerical code, and then try to evaluate the existence of additional structures in the AGN by checking their influence onto 
the resulting polarization signal and time-delays.

\begin{acknowledgements}
The authors would like to thank the anonymous referee for useful suggestions that helped to improve this paper.
This research has been supported by the French Programme National des Hautes Energies (PNHE). FM is grateful for CNES 
funding under the post-doctoral grant ``Probing the geometry and physics of active galactic nuclei with ultraviolet 
and X-ray polarized radiative transfer''. APRL acknowledge support from the CONICYT BECAS Chile grant no. 72150573.
\end{acknowledgements}

\bibliographystyle{aa}
\bibliography{biblio}

\end{document}